\documentclass[prb,twocolumn]{revtex4-2}
\usepackage{latexsym}
\usepackage{graphicx}

\usepackage{array}
\usepackage{verbatim}
\usepackage{amsmath}
\usepackage{color}
\usepackage{xcolor}
\usepackage{soul}
\usepackage{tabu}
\usepackage{multirow}
\usepackage{lipsum}
\usepackage[normalem]{ulem}
\usepackage{subfigure}
\usepackage{appendix}
\usepackage{dsfont}
\usepackage{amssymb}

\begin{document}
\title{Gradient expansion approximation of the inhomogeneous electron-gas revisited: Higher-order corrections}
  \author{Mario Benites, Angel Rosado, and Efstratios Manousakis}
  \address{
    Department  of  Physics, Florida State University}
  \begin{abstract}
    In our recently published work (Ref.~\onlinecite{benites2026})  we revisited the gradient expansion approximation (GEA) of the interacting electron gas,
    and recalculated the leading-order contribution—with respect to the Wigner-Seitz radius $r_s$—to the coefficient $B_{xc}[n]$ of the square of the gradient of the electron density in the high-density and slowly varying limits.
    That work resolved historical controversies regarding these coefficients
    and demonstrated that serious misconceptions have led to
incorrect constraints being imposed on popular functionals within the generalized gradient approximation (GGA).    
In the present paper, we extend this calculation to obtain the coefficient of the next-to-leading term, which scales as $r_s \ln(r_s)$ relative to
the leading order. 
First, we establish a systematic framework to evaluate
the integral expressions for the $b_{xc}$ coefficient of the leading term ($\sim q^2$) of the density-density response function in the long-wavelength limit ($q \to 0$)—a prerequisite for computing $B_{xc}[n]$. The significance of the calculation stems from the proof
that the coefficient of this $r_s \ln(r_s)$ term receives no corrections
from higher-order diagrammatic expressions. Consequently,
    our derived value serves as an exact, definitive
    constraint for future 
    GGA functional development;
    in the high-density slowly-varying limit, any valid functional
    must reproduce the exact constraints established in both our previous
    work and the present paper.
  \end{abstract}
  \maketitle
  \section{Introduction}
  Density functional theory (DFT) is widely used across various disciplines in science and engineering to determine the electronic structure of atoms, molecules, and
  crystalline solids. DFT is based on the
one-body Kohn-Sham equations\cite{PhysRev.140.A1133}, which
map the many-body problem to an equivalent system of $N$ non-interacting electrons in the presence of an effective external field. In addition to
the external field $V_{\mathrm{ion}}$ generated by the ions, this field contains the
Hartree term and the ``exchange-correlation'' potential $V_{\mathrm{xc}}$, which is a functional of the spatially dependent local ground-state density field
$n(\vec{r})$ produced by the collective presence of all of the
electrons of the interacting system.
The functional $V_{\mathrm{xc}}$ is universal \cite{PhysRev.140.A1133},
meaning it is material independent (i.e., independent of $V_{\mathrm{ion}}$),
which allows its determination using the many-body theory of a pure interacting electron-gas system alone (where $V_{\mathrm{ion}}=0$).
As a result, the problem of the interacting electron gas has been a focus of interest for
nearly a century \cite{PhysRev.46.1002,PhysRev.82.625,PhysRev.85.338,PhysRev.92.609,PhysRev.92.626,PhysRev.106.364,PhysRev.109.741,PhysRev.139.A796,PhysRev.136.B864,Ma-Brueckner,Geldart-Taylor1970,Geldart-Rasolt,Sham1971,PhysRevB.21.5469_Langreth_Perdew,PhysRevLett.45.566,Gross1981,Kleinman1984,Kleinman-Antoniewicz,Kleinman-Lee1988,Engel-Vosko1990,PhysRevB.54.17402,Svendsen1995,Vignale,benites2024}
and it is covered in classic many-body theory books \cite{Fetter,Mahan,Pines,Gorkov}.

A simple approximation to the $V_{\mathrm{xc}}$ functional is known as the local density approximation (LDA) (Ref.~\cite{benites2024} and references therein).
The local part of the functional is determined by means of the density dependence of the correlation energy
of the homogeneous electron gas.

Ma and Brueckner \cite{Ma-Brueckner} (MB) extended the LDA by developing the gradient-expansion approximation (GEA)\cite{PhysRevLett.77.3865,Geldart-Rasolt,PhysRevLett.100.136406-PBEsol}. The GEA is a semi-local approximation where a term, which is a functional of the density and its spatial gradient, is added to the LDA part.
This GEA contribution is determined from the response of the electron gas to a weak external spatially varying field.
In the GEA, within the weak perturbation limit, where the density is slowly varying, and in
the high-density limit, the problem translates into finding a coefficient of the term proportional to the square of the density gradient. This is achieved by calculating
the static proper-polarization function $\Pi^*(q,\omega=0)$ and extracting the
terms contributing to the coefficient $b_{xc}$  of $q^2$ in the long-wavelength
and high-density limits. The proper-polarization function is discussed further
in Sec.~\ref{Gradient_approx}, and is thoroughly defined in the literature \cite{Fetter,Mahan,Gorkov}.

When attempting to describe the electronic structure of real materials, the dimensionless quantity
\begin{equation}
  \boldsymbol{s}(\vec r\,) \equiv {{\nabla n(\vec r\,)} \over {2 k_F n(\vec r\,)}},
  \label{dimensionless_gradient}
\end{equation}
where $k_F$ is the Fermi wavevector is not necessarily small in certain regions of a given material. To capture the contribution of such
non-perturbative effects in the functional, the generalized gradient approximation (GGA) was introduced\cite{PhysRevLett.77.3865}. Within the GGA, an ad hoc
form of the
functional of the gradient of the density is implemented.
Therefore, while the validity of the GEA is restricted to the
limits of small $|\boldsymbol s|$ and small $r_s$, its importance stems from
the fact that any GGA functional must be forced to obey the constraints
imposed by the GEA in the regime of its validity. Namely,
 the coefficient $b_{xc}$ of the $q^2$ term in $\Pi^*(q, 0)$, which is related to
 the coefficient of $|\boldsymbol s|^2$, must be reproduced by any GGA in the
 small $r_s$ limit.

However, a controversy existed  prior to our
previous work\cite{benites2026} regarding the calculations of the separate
exchange
$b_x$ and correlation $b_c$ contributions to the total coefficient $b_{xc}=b_x+b_c$.   This controversy was discussed and resolved in our previous
work \cite{benites2026},
where we revisited all the previous calculations and exposed misconceptions that led to incorrect conclusions. We developed a generalized framework to
apply the necessary regularization to the singular Coulomb interaction potential. We demonstrated that the separate exchange ($b_x$) and correlation ($b_c$) contributions to the total coefficient $b_{xc}$
possess regularization-scheme dependent values even when the regulator is set to zero at the end of the calculation. This scheme-dependence implies that it is impossible to define such a separation in a meaningful way. Conversely,
we showed that when the regulator is set to zero at the end of the calculation, the combined coefficient $b_{xc}$  is regularization-scheme independent and, thus, yields a unique, unambiguous  value.  Consequently, we concluded that it is incorrect to separate those two terms when constructing a generalized-gradient-approximation (GGA) contribution to the density functional. Nevertheless, this flawed separation appears to be a common approach in most popular GGA functionals, where various constraints are applied to each contribution individually.

   In summary, in our previous work\cite{benites2026} we calculated the leading order in the $r_s$ expansion of the coefficient $b_{xc}$ and demonstrated that it
   is independent of the regularization scheme, whereas $b_x$ and $b_c$ are not.

   In the present paper, we generalize our previous approach in such a way to allow us to calculate the next-to-leading order (NLO) in $r_s$ contribution to $b_{xc}$; as we demonstrate, this NLO contribution is proportional to  $r_s \ln r_s$ relative to the previously calculated leading order (LO).
   We find that the value of the $b_{xc}$ coefficient in the
   limit of small $r_s$ is given by
   \begin{align}
  b_{xc} & = b^{(0)}_{xc} r_s^2  + b^{(1)}_{xc}  r^3_s \ln(r_s) +
  O(r_s^3) + ..., \\
    b^{(0)}_{xc}  &= 0.3164 \frac{(e m\alpha a_B)^2}{\pi^3}, \hskip 0.1 in   b^{(1)}_{xc}  = -\frac{29}{1944}\frac{e^4 (m a_B)^3}{\pi^5},
\end{align}
   where $\alpha=(4/9 \pi)^{1/3}$, $a_B$ is the Bohr radius; $m$ and $e$ are the   electron mass and charge respectively. 
We show that  no other contributions to the coefficient of this term arise
from any higher-order diagrams that have been neglected. Thus, the value of the
   coefficient of this $r_s \ln r_s$ term should be considered exact; namely,
   free for additional higher-order diagrammatic corrections.

Therefore, any consistently derived GGA functional in the limit $r_s \to 0$
must satisfy the following constraint for the coefficient of the $|\boldsymbol s|^2$ (For its definition see next section) :
\begin{equation}
B_{xc} = \frac{1}{r_s^4} \left[ 0.029116 - \frac{29}{864  \pi^3} r_s \ln r_s \right],
\label{Bxc_final}
\end{equation}
 where atomic units have been adopted in the expression above.

The paper is organized as follows. In Sec.\ref{Gradient_approx} we provide an overview of the GEA.
In Sec.~\ref{integrals_b_xc} we present the expressions for the integrals contributing to $b_{xc}$ obtained from the perturbative expansion of the
proper polarization function. In Sec.~\ref{finding_NLO} we execute the next-to-the-leading order  expansion of $b_{xc}$ in powers of  $r_s$.
In Sec.~\ref{Discussion} we summarize our calculation and present our conclusions.

\section{Review of the Gradient Expansion Approximation}
\label{Gradient_approx}
One way to improve the Local Density Approximation (LDA) is to add to it a correction, as in the generalized gradient approximation (GGA), which, in the limits of small $r_s$ and smooth electron-density variation, can be calculated by the gradient expansion approximation (GEA)\cite{benites2026}.  We worked with the following form of the exchange-correlation functional form: 
\begin{flalign}
E_{xc}[n(\vec r)] = \!\!\int \!\!d^3r' \Bigl [ A_{xc}[n({\vec r}\,')] + B_{xc}[n(\vec{r}\,')]| \boldsymbol{s}(\vec{r}\,')|^2 \Bigr ], \!\!\!&&
\label{xc_GE_functional}    
\end{flalign}
In our previous work, we only worked up to second order within the GEA in the limit of slowly varying density, where the coefficient $B_{xc}[n]$ can be found systematically by expressing the exchange-correlation energy functional in wave-vector space and obtaining the coefficient of $q^2$. Terms involving $\nabla^2 n(\vec{r}\,)$ start at order $q^4$, and we will ignore them in the present paper.
As a starting point, one must first minimize the functional $E_{xc}[n(\vec{r}\,)]$ at $n=n_0$, where $n_0$ is the homogeneous part of the electron density. After this, the coefficient $B_{xc}[n]$ can be found within linear response by studying the influence of an external potential acting on the interacting electron gas. The next step is to express the exchange-correlation functional in wave-vector space, obtaining the following form of $E_{xc}[n]$
\begin{equation}
E_{xc}[n] = E_{xc}[n^0]+\sum_{\vec{q}} K_{xc}[q,0] \delta n_{\vec{q}}\, \delta n_{-\vec{q}},
\label{Exc_inhomoegeneous_final}    
\end{equation}
where $K_{xc}[q,0]$ is given by the following expression
\begin{equation}
K_{xc}[q,0] = -\frac{1}{2}\left[\frac{1}{\Pi^{*}(q,0)}-\frac{1}{\Pi_0(q,0)}\right].
\label{B_xc_wavevector}
\end{equation}

\begin{figure*}[htp]
   \begin{center}
   \includegraphics[scale=0.37]{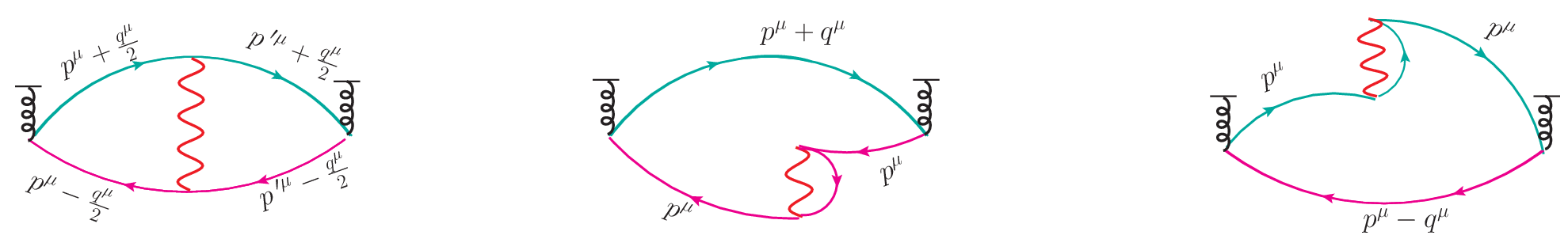} 
   \end{center}
   \caption{Diagrammatic contributions to the irreducible polarization function up to first order in an expansion of $\tilde{V}_{\lambda}(p)$ given by Eq.~\ref{yukawa2}. In our notation, $p^{\mu}=(p^0,\vec{p})$ and $q^{\mu}=(0,\vec{q})$.}
   \label{GA_Fock}
\end{figure*}

\begin{figure*}[htp]
   \begin{center}
   \includegraphics[scale=0.26]{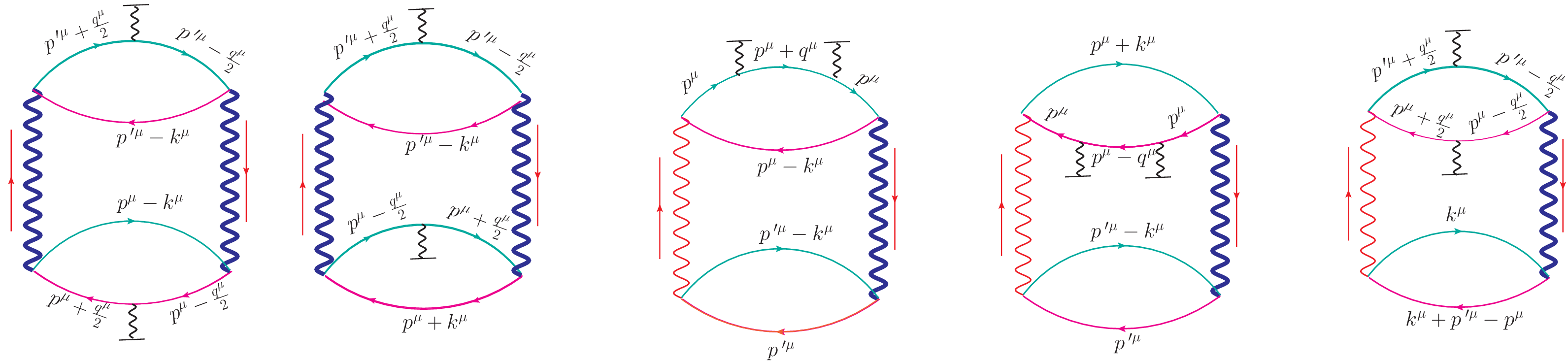} 
   \end{center}
   \caption{Remaining diagrammatic contributions to $\Pi^{xc}(q,0)$ to be combined with the diagrams in Fig.~\ref{GA_Fock}. These diagrams contribute to the same order in $r_s$ in the long-wavelength limit. The red color represents the regularized-Coulomb interaction line, while the blue color represents the RPA-renormalized interaction line. The solid green (fuchsia) lines represent the fermionic non-interacting electron (hole) propagator, while the curly line represents the insertion due to the weak external potential. The frequency and momentum variables follow the same shorthand notation as in Fig.~\ref{GA_Fock}.}
   \label{Pi_c}
\end{figure*}

In our previous work \cite{benites2026} we revisited the calculation of $b_{xc}$, where we calculated the proper-polarization function $\Pi^*(q,0)$ and performing an expansion in the $q \to 0$ limit to extract the coefficient of $q^2$; which contributes to $B_{xc}[n({\vec r})]$. In that work, we separated $\Pi^*(q,0)$ from $\Pi_0(q,0)$, which is the leading order of the density-response function $\chi(q,0)$ in the static limit. In general, the density-response function is defined as follows:
\begin{equation}
\chi (\vec{r},\vec{r}\,',t) = \langle \Psi_0|T \{\hat{n}(\vec{r},t) \hat{n}(\vec{r}\,',0) \}| \Psi_0\rangle,
\label{Chi}   
\end{equation}
where $| \Psi_0 \rangle$ is the ground-state wave function of the homogeneous interacting electron gas,
and $\Pi^*(q,0)$ can be conveniently expressed as:
\begin{equation}
\Pi^*(q,0) = \Pi_0(q,0)+\Pi^{xc}(q,0),
\label{Proper_separation}    
\end{equation}
where $\Pi^{xc}(q,0)$ is the sum of all proper-polarization functions illustrated in Figs.~\ref{GA_Fock} and~\ref{Pi_c}. In our previous work\cite{benites2026}, it was convenient to express $\Pi^{xc}(q,0)$ as the following sum:
\begin{equation}
\Pi^{xc}(q,0) = \Pi^x(q,0)+\Pi^c(q,0),    
\end{equation}
where in the $q \to 0$ limit, the expansions of $\Pi_0(q,0)$, $\Pi^{xc}(q,0)$, $\Pi^x(q,0)$, and $\Pi^c(q,0)$ are expressed as follows:
\begin{align}
\Pi_0(q,0) &= \frac{e^2 m^2}{\pi^2} \left(-\frac{1}{\alpha r_s} + \frac{\alpha r_s}{12} q^2\right), \label{polarization_proper_twiddles}\\
\Pi^{xc}(q,0) &= a_{xc}+b_{xc}q^2, \label{polarization_xc_twiddles}\\
\Pi^{x}(q,0) &= a_{x}+b_{x}q^2, \label{polarization_x_twiddles}\\
\Pi^{c}(q,0) &= a_{c}+b_{c}q^2, \label{polarization_c_twiddles}
\end{align} 
where $\Pi^x(q,0)$ is the sum of the diagrams illustrated in Fig.~\ref{GA_Fock}. Its expansion in the $q \to 0$ limit and the extraction of the $q^2$ coefficient $b_x$---which contributes to the exchange part of the density-gradient-squared term within the GEA---have been previously investigated by several authors, including Geldart-Taylor, Sham, Antoniewicz-Kleinman, Kleinman-Lee, Engel-Vosko (EV), and Svendsen and von-Barth. Likewise, $\Pi^c(q,0)$ is the sum of the proper polarization functions shown in Fig.~\ref{Pi_c}, where the $q^2$ coefficient $b_c$ was first extracted by Ma-Brueckner, and the calculation was later revisited by Geldart-Rasolt, Langreth-Perdew, and Kleinman-Tamura. As we pointed out in Ref.~\cite{benites2026}, a historical controversy arose due to disagreements among the values of $b_x$ and $b_c$ obtained by these different authors. In fact, we proved that $b_x$ and $b_c$ are individually regulator dependent because the integrals defining them do not converge; however, their combination cancels this regulator dependence, yielding a unique value for $b_{xc}=b_x+b_c$, the $q^2$ coefficient of $\Pi^{xc}(q,0)$. 

To expose the ambiguity in the values of $b_x$ and $b_c$ in the calculations of the proper polarization functions shown in Figs.~\ref{GA_Fock} and~\ref{Pi_c}, we introduced a $k_F$-dependent regulator $\lambda(k_F)$ into a Yukawa-like potential, given by the following expression in momentum space:
\begin{equation}
\tilde{V}(k) = \frac{4 \pi e^2}{k^2+\lambda^2(k_F)},
\label{yukawa2}    
\end{equation}
where we have set $\lambda(k_F)=\lambda_c \beta(k_F)$, with $\lambda_c$ being a coefficient whose limit $\lambda_c \to 0$ is taken at the very end of the calculation. This $k_F$-dependent regulator will also be used in the present work. However, as we discuss in the next section, for some terms contributing to the integral expressions of $\Pi^{xc}(q,0)$ that contain the RPA-screened interaction, where the Lindhard function acts as an emergent regulator, it may not be necessary to retain a non-zero regulator $\lambda(k_F)$ during the integral evaluations. The RPA-screened interaction is given by:
\begin{align}
\tilde{V}_{e}(k^{\mu}) &= \frac{\tilde{V}(k)}{\epsilon(k^{\mu})},
\label{V_RPA}\\
\epsilon(k^{\mu}) &= 1-\tilde{V}(k)\Pi_0(k^{\mu}),
\end{align}
where $k^{\mu}=(k^0, \vec{k}\,)$ is a shorthand notation to indicate frequency and momentum dependence.
In general, $\Pi_0(k^{\mu})$ is defined by the following integral:
\begin{equation}
\Pi_0(k^{\mu}) = -2i\int\frac{d^4p}{(2\pi)^4} G^0(p^{\mu}+k^{\mu}) G^0(p^{\mu}),
\label{Pi_0}
\end{equation}
where $G^0(p^{\mu})$ is the non-interacting Green's function given by the following expression:
\begin{equation}
G^0(p^{\mu}) = \frac{1}{p^0-\epsilon^0_p+\mu_0+i\eta\,\mathrm{sign}(\epsilon^0_p - \mu_0)},
\label{Green_function}
\end{equation}
and $\epsilon^0_p$ and $\mu_0$ are the non-interacting energy dispersion and chemical potential, respectively.

The coefficients $a_0$, $b_0$, $a_{xc}$, and $b_{xc}$ are determined within the GEA, which allows us to find the coefficient $B_{xc}[n]$ of the gradient of the electron density squared in the exchange-correlation functional $E_{xc}[n]$. In particular, $a_0$ and $b_0$ are the leading term and $q^2$ coefficient in the long-wavelength ($q \to 0$) expansion of $\Pi_0(q,0)$ given by Eq.~\ref{polarization_proper_twiddles}, respectively. The expression for $B_{xc}[n]$ in terms of $a_0$, $b_0$, $a_{xc}$, and $b_{xc}$ is obtained by expanding the proper polarization functions, given by Eqs.~\ref{polarization_proper_twiddles}--\ref{polarization_c_twiddles}, in the $q \to 0$ limit. This expression for $B_{xc}[n]$ is given by the following
\begin{align}
  B_{xc}[n] &= (2 k_F n)^2 K^{\prime\prime}_{xc}, \label{B_coeff}    
\\
K^{\prime\prime}_{xc} &\equiv \left.\frac{\partial^2 K_{xc}}{\partial q^2}\right|_{q=0} = \frac{b_{xc}a_0-2b_{0}a_{xc}}{2 a^3_0}.
\end{align}


In our recent work \cite{benites2026}, we found that the coefficients $b_x$ and $b_c$ contribute at leading order in the Wigner-Seitz radius as $r_s^2$. We reported these values of the coefficients in Table~\ref{table_exchange_correlation} for different choices of the regulator $\lambda(k_F)\neq 0$. While different regulators yield distinct individual values for $b_x$ and $b_c$, their combination $b_{xc}$ is always guaranteed to remain unique.  
\setlength{\tabcolsep}{4pt}
\def\arraystretch{1.5}
\begin{table}[ht]
    \centering
    \begin{tabular}{|c|c|c|c|c|c|c}
        \hline
         & Case 1 & Case 2  & Case 3 & Case 4 & Case 5 \\
        \hline
        $b_x$ & $5/72$ & $5/24$ & $5/36$ & $1/36$ &  Diverges \\
        \hline
        $b^{'}_c$ & $5/72$ & $-5/72$ & $0$ & $1/9$ & Do~not~exist\\
        \hline
        $b^{''}_c$ & $0.10359$ & $0.10359$ & $0.10359$ & $0.10359$ & Do~not~exist\\
        \hline
        $b^{'''}_{c}$ & $0.07392$ & $0.07392$ & $0.07392$ & $0.07392$ & $0.07392$ \\
        \hline
        $b_{xc}$ & $0.3164$ & $0.3164$ & $0.3164$ & $0.3164$ &  undetermined\\
        \hline
    \end{tabular}
    \caption{Leading coefficients of the $q^2$ terms corresponding to $\Pi^{xc}(q,0)$ found from the sum of the diagrams illustrated in Fig.~\ref{GA_Fock} and Fig.~\ref{Pi_c}. This table was extracted from Ref.~\cite{benites2026} and the values in the table are given in units of a common factor of
      $\tau = e^2m^2/\pi^3k_F^2$.}
    \label{table_exchange_correlation}
\end{table} 

The coefficient we found corresponds to $E[n(\vec{r}\,)]$ when we re-express Eq.~\ref{xc_GE_functional} in terms of $s^2$ instead of $|\nabla n(\vec{r})|^2$ \cite{benites2026}. Consistently with the conventions used in this work, the leading-order term in $r_s$ of $B_{xc}[n]$, which serves as a constraint in the construction of density functionals, is given precisely by the sum of Sham's value for $B_x$ and Ma-Brueckner's value for $B_c$. We have:
\begin{equation}
B^{MB-S}_{xc}[n] = \frac{0.029116}{r_s^4},
\label{B_xc_MBS}    
\end{equation}
where the value is reported in atomic units and the superscript denotes the authors who first derived the values of $B_x$ and $B_c$, with $B_{xc}=B_x+B_c$. In the next section, we summarize the calculations used to obtain the integral expressions for $b_{xc}$ in our recent paper, where we employed the $k_F$-dependent regulator $\lambda(k_F)$.

\section{The integral expressions of $b_{xc}$}
\label{integrals_b_xc}
In this section, we present the integral expressions for the proper polarization functions that contribute to $\Pi^{xc}(q,0)$ and summarize the steps taken in Ref.~\cite{benites2026} to obtain the two-dimensional integral expressions for $b_{xc}$ at any value of $r_s$ in the static, long-wavelength limit. The sum of the proper polarization functions illustrated in Figs.~\ref{GA_Fock} and~\ref{Pi_c} can be written as
\begin{equation}
\Pi^{xc}(q,0) = \sum_{i=1}^{3} \Pi^{xc}_i(q),
\label{Pi_xc_reorganized}    
\end{equation}
where $\Pi^{xc}_1(q,0)$ is the sum of the vertex bubble on the left side of Fig.~\ref{GA_Fock} and the last term on the right of Fig.~\ref{Pi_c}. $\Pi^{xc}_2(q,0)$ is the sum of the remaining diagrams from Fig.~\ref{GA_Fock} with the third and fourth diagram in Fig.~\ref{Pi_c} (counting from left to right). We label the sum of the final two diagrams in Fig.~\ref{Pi_c} as $\Pi^{xc}_3(q,0)$. The integral expressions for these proper polarization functions $\Pi^{xc}_i(q,0)$, (for $i=1,2,3$) were provided in Sec.~III of Ref.~\cite{benites2026}.

Performing a Taylor expansion of these proper polarization functions in the $q \to 0$ limit, we found that their expansion takes the following form
\begin{equation}
\Pi^{xc}_{i}(q) \approx  a^i_{xc}+b^i_{xc}q^2, (\mathrm{for} \ i=1,2,3),
\label{decomposition_pixc_i}
\end{equation}
where the integral expressions obtained for the $b^i_{xc}$ coefficients were also provided in Sec.~III of Ref.~\cite{benites2026}. We found that these $q^2$ coefficients can be conveniently recombined as a sum of four main terms: $b'_{xc}$, $b^{''}_{xc}$, $b^{'''}_{c}$ and a term that depends on the value of the GW self-energy at $p=k_F$. We have the following
\begin{equation}
\sum_{i=1}^3 b^i_{xc} = b^{'}_{xc}+b^{''}_{xc}+b^{'''}_{c}-\frac{m^2 \Sigma_{GW}(k_F,0)}{12 \pi^2 k_F^3},
\label{b_xc_sum}
\end{equation}
where the GW self-energy is defined as follows
\begin{align}
\Sigma_{GW}(p^{\mu}) &= \Sigma_{F}(p^{\mu})+\Sigma_r(p^{\mu}),\label{self-energy_GW}\\
\Sigma_{F}(p^{\mu}) &\equiv i\int \frac{d^4p'}{(2\pi)^4} e^{ip'^0 \eta} \tilde{V}(\vec{p'}-\vec{p}) G^0(p',p'^0),
\label{self-energy_Fock}\\
\Sigma_r(p^{\mu}) &= -i\int\frac{d^4p'}{(2\pi)^4}\tilde{V}_i(\kappa^{\mu}) G^0(p'^{\mu}),
\label{self-energy_ring}   
\end{align}
where
\begin{equation}
  \tilde{V}_i(\kappa^{\mu}) \equiv
  \tilde{V}(\kappa^{\mu})-\tilde{V}_e(\vec{\kappa}),
  \end{equation}
and $\kappa^{\mu} = p^{\prime \mu}-p^{\mu}$, where $p^{\mu}=(p^0,\vec{p}\,)$ is a shorthand notation used throughout this work, and $\Sigma_r(p^{\mu})$ is what we call the ring-like self-energy (the sole term carrying the renormalized potential $\tilde{V}_e(k^{\mu})$ within the RPA). This term is responsible for yielding higher-order terms in $r_s$ to the $b_{xc}$ coefficient in the high-density limit ($r_s \to 0$). Furthermore, it proved convenient to split the $b'_{xc}$ and $b''_{xc}$ coefficients into two contributions arising from the decomposition of the GW self-energy $\Sigma_{GW}(p^{\mu})$ and the GW vertex function $\Lambda^{GW}_2(p^{\mu})$. We obtain
\begin{equation}
b'_{xc} = b^{'}_x+b^{'}_c, \quad b^{''}_{xc} = b^{''}_x + b^{''}_c,
\label{bxc}   
\end{equation}
where the $b$-primed coefficients are given by
\begin{align}
b^{'}_x &= b^{'}_{x,1}+b^{'}_{x,2}, \label{b'_x}\\
b^{'}_{x,1} &= \frac{i}{m} \frac{\partial}{\partial \mu_0} \int d[{\bf p}] \Sigma_F(p^{\mu}) \frac{1}{2}\frac{\partial^2 G^0(p^{\mu})}{\partial \mu^2_0}, \label{b'_x1} \\
b^{'}_{x,2} &= -\frac{i}{m}\frac{\partial}{\partial \mu_0} \int d[{\bf p}] \Sigma_F(p^{\mu}) \frac{\epsilon^0_p}{9}\frac{\partial^3 G^0(p^{\mu})}{\partial \mu^3_0}, \label{b'_x2} \\
b^{''}_x &= -\frac{i}{6m} \int d[{\bf p}] \Sigma_F(p^{\mu}) \frac{\partial^3 G^0(p^{\mu})}{\partial \mu^3_0}, \label{b''_x} \\
b^{'}_c &= \frac{i}{m} \frac{\partial}{\partial \mu_0} \!\int \!d[{\bf k}] \tilde{V}_{i}(k^{\mu}) \left(\frac{1}{2}I_1(k^{\mu}) -\frac{1}{9}I_3(k^{\mu}) \right), \label{b'_c} \\
b^{''}_c &= -\frac{i}{6m} \int d[{\bf k}] \tilde{V}_{i}(k^{\mu})I_2(k^{\mu}), \label{b''_c}    
\end{align}
where $m$ is the electron mass, the integration over the bracket term represents $d[{\bf p}] \equiv d^4p/(2\pi)^4$, and the last expression (which is not split into two terms) is given by
%
 \begin{equation}
  b^{'''}_{c} = \frac{i}{24}\int \!d[{\bf k}] \left(\!\frac{\partial \Pi_0(k^{\mu})}{\partial \mu_0} \!\right)^{\!\!2} \left[\tilde{V}_{e}\nabla^2_k \tilde{V}_{e}-\left(\frac{d\tilde{V}_{e}}{dk} \right)^{\!\!2} \right],
\label{b'''_xc}
 \end{equation}
%
This last expression agrees with the $b^{'''}$ coefficient obtained by MB in Ref.~\cite{Ma-Brueckner}. The functions $I_i(k^{\mu})$ ($i=1,2,3$) are given by the following expressions
\begin{align}
I_1(k^{\mu}) = -i\int d[{\bf p}]G^0(p^{\mu}+k^{\mu}) \frac{\partial^2}{\partial \mu_0^2} G^0(p^{\mu}), \label{I1}\\    
I_2(k^{\mu}) = -i\int d[{\bf p}]G^0(p^{\mu}+k^{\mu}) \frac{\partial^3}{\partial \mu_0^3} G^0(p^{\mu}), \label{I2}\\
I_3(k^{\mu}) = -i\int d[{\bf p}]G^0(p^{\mu}+k^{\mu}) \frac{\partial^3}{\partial \mu_0^3} G^0(p^{\mu}) \epsilon^0_{\vec{p}}. \label{I3}    
\end{align}
These expressions, the $b_x$-primed and $b_c$-primed coefficients, together are contributions of $b_x$ and $b_c$, which are obtained from the separation of terms done in Eq.~\ref{b_xc_sum}. We have
\begin{align}
b_x &= b'_x+b^{''}_x-\frac{\Sigma_F(k_F,0)m^2}{12 \pi^2 k^3_F},\label{b_x_def}\\ 
b_c &= b'_c+b^{''}_c+b^{'''}_{c}+b^r_c,\label{b_c_def}
\end{align}
where we have defined the coefficient $b^r_c$ as follows
\begin{equation}
b^r_c = -\frac{\Sigma_r(k_F,0)m^2}{12 \pi^2 k_F^3},
\label{br_c} 
\end{equation}
where both self-energies were defined in Eqs.~\ref{self-energy_Fock} and~\ref{self-energy_ring}.

\begin{figure*}[htp]
   \begin{center}
   \includegraphics[scale=0.37]{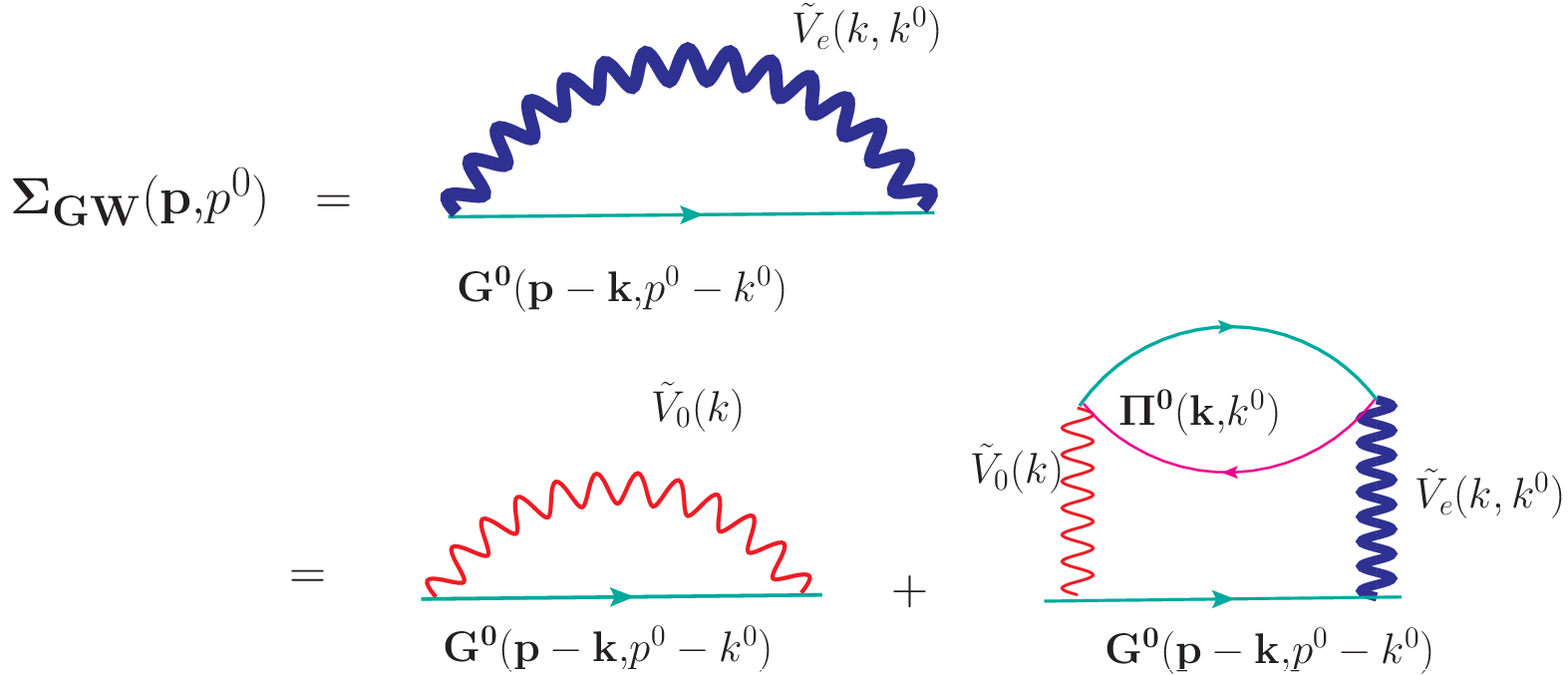} 
   \end{center}
   \caption{Diagrammatic representation of the GW self-energy. The first diagram corresponds to the Fock self-energy $\Sigma_F(p^{\mu})$ diagram and the second diagram is the ring-like self-energy diagram $\Sigma_r(p^{\mu})$.}
   \label{self_GW_fig}
\end{figure*}

We used the regularized potential $\tilde{V}(k)$ with the $\lambda(k_F)$ regulator given by Eq.~\ref{yukawa2}, where in our latest work the infinitesimal parameter $\lambda_c$  in the expression of $\lambda(k_F)$ has to be taken to $\lambda_c \to 0$ at the end of the calculation \cite{benites2026}. We proved that the integral expressions for the $b_x$-primed and $b_c$-primed coefficients can be reduced to a more compact form. For the $b_x$-primed coefficients, evaluating the frequency integrals in Eqs.~\ref{b'_x1},~\ref{b'_x2}, and~\ref{b''_x} is straightforward, while the momentum integrations are simple to calculate due to the Dirac delta functions $\delta(\epsilon^0_p-\mu^0)$ arising from the partial derivatives of $G^0(p^{\mu})$ with respect to $\mu_0$. The $b_x$ coefficient contributes only at order $r^2_s$, and we have demonstrated that it is regulator-dependent. For the $b_c$ coefficients, we reduced the dimensionality of their respective expressions by analyzing the $I_i(k^{\mu})$ integrals and the complex poles associated with the frequency component $k^0$ (see Ref.~\cite{benites2026} for details of this derivation). The resulting final expressions that we found for $b_x$, $b^{\prime}_c$, $b^{\prime \prime}_c$, and $b^{\prime \prime \prime}_c$ are given by the following
%
\begin{align}
&b^{(\lambda_c \to 0)}_x = \frac{5 m^2 e^2}{72 \pi^3 k_F^2}\left[1+2k_F\frac{\partial}{\partial k_F}\ln(\beta(k_F))\right],
\label{b_x_final}\\
&b^{'}_c = \frac{\partial}{\partial \mu_0} \int^{\infty}_0dx \int^{\infty}_0 dy Z_{\lambda'}(x,y) C(x,y),\label{b'_c_2d}\\
&b^{''}_c = \frac{m }{3 k_F^2}\int_0^{\infty} dx \int_{0}^{\infty}dy Z_{\lambda'}(x,y)I'_2(x,y),\label{b''_c_2d} \\
&\begin{aligned}[b]
b^{'''}_{c}=& \frac{m^3 e^4}{96 \pi^5 k_F^3}\int^{\infty}_0 \frac{dx}{x}\int_0^{\infty} \!\!\!dy \left(\frac{g(x,y)}{\overline{\epsilon_{\lambda}}(x,y)}\!\right)^{\!2} \times \\
&\left[1-\frac{\alpha r_s M_9(x,y)}{2 \pi \overline{\epsilon_{\lambda}}(x,y)}-\frac{(\alpha r_s M_{10}(x,y))^2}{8 \pi^2 (\overline{\epsilon_{\lambda}}(x,y))^2} \right], \label{b'''_c_2d}
\end{aligned}
\\
&Z_{\lambda'}(x,y) \equiv -{{2e^2m}\over {\pi^4}} \frac{x}{x^2+{\frac{\lambda'^2}{4}}} \left(\frac{x^2+\frac{\lambda'^2}{4}}{\overline{\epsilon_{\lambda}}(x,y)}-1 \right), \label{Z_function}
\end{align}
%
where the symbol $\lambda'=\lambda_c\beta(k_F)/k_F$ is used as a reminder that the function is dependent on $\lambda$, $x=q/2k_F$ and $y=m \nu/k_F q$. The superscript ($\lambda_c \to 0$) indicates that the regulator $\lambda_c$ is treated as a constant throughout the integration and taken to zero $\lambda_c \rightarrow 0$ at the very end of the calculation. The function $C(x,y)$ is given by
\begin{align}
&\!C(x,y) \!=\! \frac{M_5(x,y)}{D(x,y)}\!+\!\frac{M_6(x,y)}{9D(x,y)^2}\!-\!\frac{g(x,y)}{24 x} \!+\! \frac{I'_2(x,y)}{9}, \label{C_xy}\\
&M_5(x,y) = \left(\frac{1}{6}\!-\!\frac{5x^2}{18}\right)\!(1\!+\!y^2\!-\!x^2)\!+\!\frac{y^2}{9}(2\!+\!5x^2), \label{M5}\\
&M_{6}(x,y) = y^2(x^4+2x^2y^2+y^4-1), \label{M6}\\
&M_{10}(x,y) = \frac{4-3 Q(x,y)}{2}-\left(\frac{y^2+x^2}{2x}\right)g(x,y), \label{M10}\\
&M_{8,2} = \frac{(x^2+y^2)(x^2-3y^2-1)}{x^2 D(x,y)}, \label{M82} \\
&\begin{aligned}[b]
M_9(x,y) =& 4-\left(\frac{7y^2+3x^2}{4x}\right)g(x,y)-\frac{3}{2}Q(x,y) \\
& -x^2M_{8,2}, \label{M9} \\
\end{aligned}
\end{align}
where $Q(x,y)$ is the Lindhard function for imaginary frequency, given by the following expression
\begin{align}
Q(x,y) &= 2 +\left( \frac{1+y^2-x^2}{2x} \right)g(x,y)-g_t(x,y), \label{Q function}\\
g_t(x,y) &= 2y\left[\tan^{\!-1}\!\left(\frac{1+x}{y}\right)+\tan^{\!-1}\!\left(\frac{1-x}{y}\right)\right], \label{g2}
\end{align}
the functions $D(x,y)$ and $g(x,y)$ are given by
\begin{align}
D(x,y) &= \left((1+x)^2+y^2\right)\left((1-x)^2+y^2\right), \label{D}\\
g(x,y) &= \ln \left|\frac{(1+x)^2+y^2}{(1-x)^2+y^2} \right|,\label{g(x,y)}
\end{align}
and the function $I'_2(x,y)$ is given by the following expression
\begin{equation}
I'_2(x,y) = \frac{M_1(x,y)}{D(x,y)}+\frac{M_2(x,y)}{D(x,y)^2},
\label{I'_2}     
\end{equation}
where the functions $M_i(x,y)$ (for $i=1,2,3,4$) are
\begin{align}
M_1(x,y) &= x^2y^2-\frac{1}{4}(y^2+1-x^2)(3+2x^2),\label{M1}\\    
M_2(x,y) &= \left(\frac{3}{4}-x^2\right)N_2(x,y)+N_3(x,y),\label{M2}\\
N_2(x,y) &= (y^2+1-x^2)^2-4x^2y^2,\label{N2}\\
N_3(x,y) &= 4y^2(y^2+1-x^2)\left(\frac{3}{4}+x^2 \right).\label{N3}     
\end{align}

The function $\overline{\epsilon_{\lambda}}(x,y)$ arises from the the denominator of ${\tilde V}(k)$ multiplied by the dielectric function, after factoring out $4k_F^2$ following the transformation of momentum and frequency variables to $x$ and $y$. We obtain
\begin{equation}
\overline{\epsilon_{\lambda}}(x,y) = x^2+\frac{\lambda'^2}{4}+\frac{\alpha r_s Q(x,y)}{4 \pi}.
\label{epsilon_bar}    
\end{equation}

The final results for the leading-order (in $r_s$) contributions from $b^{'}_c$, $b^{''}_c$ and $b^{'''}_{c}$ obtained in Ref.~\cite{benites2026} are given by
%
\begin{align}
b^{'(\mathrm{LO})}_{c} &= \frac{5 e^2 m^2}{72\pi^3 k_F^2} \left[1-2k_F\frac{\partial}{\partial k_F}\ln(\beta(k_F)) \right],\label{b'_c_final_result}\\
b^{''(\mathrm{LO})}_{c} &= 0.82872\frac{e^2 m^2 }{(2 \pi)^3 k_F^2},\label{b''_c_final_result}\\
b^{'''(\mathrm{LO})}_{c} &=  0.59136 \frac{e^2 m^2}{(2 \pi)^3 k_F^2},\label{b'''_c_final_result}    
\end{align}
%
where the superscript ``LO" denotes the lowest order in $r_s$.
These final expressions for the leading-order terms in $r_s$ of both the $b_c$-primed and $b^{'''}_{c}$ coefficients were obtained in our previous work \cite{benites2026} by focusing on the contribution from the integration region $x \in (0,1/2)$. This interval corresponds to wavevector integrations over $k \in (0,k_F)$. The primary goal of the present work is to provide a deeper explanation of why this specific region is crucial for capturing not only the leading-order terms in $r_s$, but also the next-to-leading order (NLO) contributions.

As demonstrated later in this work, the $b'_c$, $b^{''}_c$, $b^{'''}_{c}$ coefficients, along with the self-energy term $\Sigma_r(k_F,0)$ in $b^r_c$, contribute to the NLO in the $r_s$ expansion for $b_{xc}$, as we will explain in the next section.

\section{Next-to-the-leading order in the $r_s$ expansion of $b_{xc}$}
\label{finding_NLO}
Since the integrals entering $b_x$ only yield $r^2_s$ contributions, our focus turns to the coefficients that contribute to $b_c$, as seen in Eq.~\ref{b_c_def}. Furthermore, we have ruled out any contributions to the NLO in $r_s$ to $b_{xc}$ arising from any other proper polarization functions than those illustrated in Figs.~\ref{GA_Fock} and~\ref{Pi_c}. Starting from the expression in Eq.~\ref{b_c_def}, we present a systematic method to evaluate every term in the $r_s$ expansion for the $b_c$-primed coefficients, $b^{\prime \prime \prime}_c$, and the self-energy value $\Sigma_r(k_F,0)$ that contributes to $b_c$. Finally, we show why no additional NLO contributions in $r_s$ originate from any other proper polarization diagrams. 

\subsection{The systematic method to extract every coefficient of the $r_s$ expansion of $b_c$}
The systematic approach presented in this work, which finds the NLO terms in $r_s$ of every term contributing to $b_c$ in Eq.~\ref{b_c_def}, consists of four main steps. 

Step 1 consists of checking whether the integral expressions from these terms can be split into a sum of distinct integrals. Specifically, we separate those terms where setting $\lambda_c=0$ directly in the integrand yields no difference from those where such a substitution cannot be made prior to integration. If this separation is achievable, we first evaluate the integrals for which taking $\lambda_c\to0$ is valid only after completing the integration. This subtle distinction is relevant solely for the $b^{'}_c$ and $b^{''}_c$ coefficients.

Step 2 consists of splitting the domain of integration for the remaining two-dimensional integrals, specifically those where setting $\lambda=0$ directly in the integrand is permissible prior to integration. The integration domain is divided into Region 1, $x \in (0,1/2]$, and Region 2, $x \in (1/2,\infty)$. More generally, any cutoff choice $x=x_c$ with $x=1/2$ could be used to split the domain without altering the resulting coefficients in the $r_s$-expansion. Our specific choice of $x_c=1/2$ for Region 1 corresponds to the wavevector interval $k \in (0,k_F)$, which captures the long-wavelength infrared behavior of the Coulomb interaction. In Region 1, we found that the general expression that we obtained has the following structure
\begin{equation}
L^{1}_{im}(r_s) \equiv \int^{\infty}_0 dy \int^{\frac{1}{2}}_0 dx \frac{f_i(x,y)}{[\bar{\epsilon}(x,y)]^m},
\label{L1_im_structures}
\end{equation}
where $i=1,\dots,9$, $m \geq 1$, and we have dropped the $\lambda$ subscript at the integrand level, since we set $\lambda=0$ prior to integration. The family of functions $f_i(x,y)$ corresponds to each integral expression that we would obtain for $b^{\prime}_c$ ($i=1$), $b^{\prime \prime}_c$ ($i=2$), $\Sigma^{\lambda \to 0}_r(k_F,0)$ ($i=3$), and $b^{\prime \prime \prime}_{c}$ ($i=4,5,6,7,8,9$), as will be shown later in this work. Furthermore, the functions $f_i(x,y)$ are odd in the variable $x$ arising from the separation mentioned in the previous step, if applicable. Similarly, Region 2 can be expressed in general as the following two-dimensional integral
\begin{equation}
L^{2}_{im}(r_s) \equiv \int^{\infty}_0 dy \int^{\infty}_{\frac{1}{2}} dx \frac{f_i(x,y)}{\left[\bar{\epsilon}(x,y)\right]^m},
\label{L2}    
\end{equation}
where in this work we are only interested in the case $m\geq 1$.

We find that NLO $r_s$ terms originate exclusively from Region 1 and can be extracted systematically via a Taylor expansion of $f_i(x,y)$ in the limit $x \to 0$. Given the odd symmetry $f_i(x,y)=-f_i(-x,y)$, its Taylor expansion takes the following form
\begin{equation}
f_i(x,y) = \sum^{\infty}_{n=0} \frac{x^{2n+1}}{(2n+1)!}f^{(2n+1,0)}_i(0,y),
\label{f_taylor}    
\end{equation}
where the superscript denotes crossed partial derivatives, $f^{(i,j)}(x,y)=\partial^i_x \partial^j_y f(x,y)$. Additionally, we perform a Taylor expansion of $Q(x,y)$ in the limit $x \to 0$, and it is convenient to re-express $\bar{\epsilon}(x,y)$ as follows
\begin{align}
\bar{\epsilon}(x,y) &= \Delta_1(x,y)+\frac{\alpha r_s}{4 \pi}\Delta_2(x,y),
\label{epsilon_bar_reexpress}\\
\Delta_1(x,y) &= x^2\gamma(r_s,y)+\frac{\alpha r_s}{4 \pi}Q(0,y),\label{Delta1}\\
\Delta_2(x,y) &= \sum^{\infty}_{n=2} \frac{x^{2n}}{(2n)!}Q^{(2n,0)}(0,y),\label{Delta2}\\
Q(x,y) &= Q(0,y)+\frac{x^2}{2}Q^{(2,0)}(0,y)+\Delta_2(x,y),\label{Q_taylor_expansion}
\end{align}
where $\Delta_2(x,y)$ represents the Taylor expansion of $Q(x,y)$ in the limit $x \to 0$, excluding the leading and NLO order terms. The prefactor $\gamma(r_s,y)$ yields partial contributions to the coefficients of the $r_s$ terms originating from $b_c$ by using Eq.~\ref{b_c_def}, and is given by
\begin{equation}
\gamma(r_s,y) = 1+\frac{\alpha r_s}{8 \pi} Q^{(2,0)}(0,y).
    \label{gamma_rs_y}
\end{equation}
Notice that the Taylor expansion on $Q(x,y)$ given by Eq.~\ref{gamma_rs_y} contains only powers of the form $x^{2n}$ ($n \in \mathds{Z}^+$), since this function is even in the variable $x$.

Next, we factor out $\Delta_1(x,y)$ from $\bar{\epsilon}(x,y)$ and expand the remaining factor as a geometric series, yielding
\begin{equation}
\frac{1}{\bar{\epsilon}(x,y)^m} \!\!\stackrel{x \to 0}{=}\! \sum^{\infty}_{l=0} \left(\!-\frac{\alpha r_s}{4 \pi} \!\right)^{\!l}  \!\frac{(l\!+\!m\!-\!1)!}{l!(m\!-\!1)!}  \frac{[\Delta_2(x,y)]^l}{[\Delta_1(x,y)]^{l+m}}.\\
\label{denominator_epsilon_expanded}    
\end{equation}
Substituting this expression along with Eq.~\ref{f_taylor} into the integral in Eq.~\ref{L1_im_structures}. We obtain the following
\begin{align}
&L^1_{im}(r_s) \!=\!\! \sum^{\infty}_{n,l=0}\!\!\left(-\frac{\alpha r_s}{4 \pi} \right)^{\!l}\!\!\!\int^{\infty}_0 \!\!\!\!\!dy \,\Gamma^{2n+1}_{lm}(r_s,y)\frac{f^{(2n+1,0)}_i(0,y)}{(2n+1)!}, 
\label{L_im_structures_2}\\
&\Gamma^{n}_{lm}(r_s,y) \equiv \frac{(l+m-1)!}{l!(m-1)!} \int^{\frac{1}{2}}_0 dx \frac{x^{n}[\Delta_2(x,y)]^l}{[\Delta_1(x,y)]^{l+m}},\label{Gamma_n_l_m_def}
\end{align}
where $m\geq 1$. Here, the NLO contributions in $r_s$ to $b^{\prime}_c$, $b^{\prime \prime}_c$, $b^{\prime \prime \prime}_{c}$ and $\Sigma^{\lambda \to 0}_r(k_F,0)$ arise solely from $\Gamma^{2n+1}_{0m}(r_s,y)$, as the terms with $l\geq 1$ yield higher-order contributions in $r_s$, as shown later in this work. In particular, the NLO terms in $r_s$ originating from $L_{im}$ can be written as
\begin{equation}
L^1_{im}(r_s) \approx \sum^{\infty}_{n=0} \int^{\infty}_0 dy \Gamma^{2n+1}_{0m}(r_s,y)\frac{f^{(2n+1,0)}_i(0,y)}{(2n+1)!}. 
\label{L1_im_l=0}    
\end{equation}

In Region 2, we exploit the property that $Q(x,y) \to 0$ for $x > 1/2$ across all $y \in \mathbb{R}$. Focusing on the $r_s \to 0$ limit allows us to Taylor expand the denominator in the integrand of Eq.~\ref{L2} as
\begin{equation}
\frac{1}{[\bar{\epsilon}(x,y)]^m} = \sum^{\infty}_{l=0} \frac{(l+m-1)!}{l!(m-1)!}\left(-\frac{\alpha r_s}{4 \pi}\right)^{\!l} \frac{[Q(x,y)]^l}{x^{2(l+m)}}, 
\label{denominator_epsilon_expanded_large_x}    
\end{equation}
where $x > 1/2$. Substituting this expansion into Eq.~\ref{L2}, we obtain its expansion in $r_s$. We have
\begin{align}
L^{2}_{im}(r_s) =& \sum^{\infty}_{l=0}\left(-\frac{\alpha r_s}{4 \pi} \right)^l\int^{\infty}_0 \!\!\!dy \int^{\infty}_{\frac{1}{2}} \!\!\!dx \frac{(l+m-1)!}{l!(m-1)!}\nonumber \\
&\times \frac{f_i(x,y)[Q(x,y)]^l}{x^{2(l+m)}},
\label{L2_im_expansion_rs}
\end{align}
which yields contributions in $r_s$ only of the form $r_s^n$ ($n \in \mathds{Z}^+$).

Step 3 of our systematic method consists of expressing the integrals that contribute to $b^{\prime}_c$, $b^{\prime \prime}_c$, $b^{\prime \prime \prime}_{c}$ and $\Sigma_r(k_F,0)$ in terms of the integral kinds $L^1_{im}(r_s)$ and $L^2_{im}(r_s)$. In this step, the integral $L^1_{im}(r_s)$ decomposes into the other integral kinds $\Gamma^n_{lm}(r_s,y)$, where the NLO term is extracted exclusively at $l=0$. Higher-order terms in $r_s$ beyond NLO originate from the contributions with $l\geq0$, as well as from $L^2_{im}(r_s)$ for any $m \in \mathbb{Z}^+$. 

In Appendix~\ref{Gamma_integral_kinds_section}, we discuss the integral kinds $\Gamma^{2n+1}_{lm}(r_s,y)$ in greater depth, expanding them in the $r_s \to 0$ limit and retaining terms only up to NLO, in line with the primary goal of this work. In Appendix~\ref{different_contribution_types_appendix}, we address how $\Gamma^{2n+1}_{lm}(r_s,y)$ gives rise to two distinct types of contributions (denoted as $\mathcal{D}$-type and $\mathcal{H}$-type) to the coefficients of the $r_s$ terms of the $b_c$-primed terms and $\Sigma_r(k_F,0)$. The $\mathcal{D}$-type contributions arise by Taylor expanding $f(x,y)$ as $x \to 0$ while setting $\bar{\epsilon}(x,y) \approx \Delta_1(x,y)$ and $\gamma(r_s,y) \to 1$ at the denominator. The $\mathcal{H}$-type contributions encompass the remaining higher-order $r_s$ corrections omitted from the D-type evaluation, which can still contribute to the NLO coefficient under certain scenarios. In the next section, we apply our systematic method to obtain the coefficient of the NLO $r_s$ term from all quantities contributing to $b_c$ in Eq.~\ref{b_c_def}. 

\subsection{Application of our systematic method for every term of $b_c$}
We take the first step of our systematic approach, which is to separate the terms of the integral expressions of one of the quantities contributing to $b_c$ in Eq.~\ref{b_c_def}. The only terms that this step applies to are $b^{\prime}_c$ and $b^{\prime \prime}_{c}$, since $\Sigma^{\lambda}_r(k_F,0)$ and $b^{'''}_{c}$ will not change the value of the integral if set $\lambda=0$ before performing the integration. The reason for this is that the term $\alpha r_s Q(x,y)/4\pi$ acts as a regulator for the integration in the denominator term $\bar{\epsilon}(x,y)$ at the integration level. By using Eqs.~\ref{b'_c_2d}--\ref{Z_function}, we have proven in our previous work \cite{benites2026} that the coefficient can be expressed as follows
\begin{flalign}
b^{'}_c =& \sum^3_{i=1} b^{\prime}_{ci},\label{bc_prime_sum}\\
b^{'}_{c1} =& -\frac{e^4m^3}{2 \pi^5 k^3_F} \int^{\infty}_0 dy \int^{\infty}_0 dx \frac{x C(x,y)Q(x,y)}{[\bar{\epsilon}_{\lambda}(x,y)]^2},\label{b'c1}\\
b^{\prime}_{c2} =& \frac{e^4 m^3}{4 \pi^5 k_F^3 } \int^{\infty}_0 \!\!dy \int^{\infty}_0 \!\!dx \frac{\lambda'^2 x C(x,y) Q(x,y)}{[x^2+(\frac{\lambda'}{2})^2][\bar{\epsilon}_{\lambda}(x,y)]^2} \nonumber \\
&\times\left(1-k_F\frac{\partial \ln(\beta(k_F))}{\partial k_F}\right),\label{b'_c2}
\end{flalign}
\begin{flalign}
b^{\prime}_{c3} =&
-\frac{e^2 m^2}{\pi^4 k_F^2} \int^{\infty}_0 \!\!\!dy \int^{\infty}_0 \!\!\!dx \frac{\lambda'^2 x C(x,y)}{[x^2+(\frac{\lambda'}{2})^2]^2} \nonumber \\
& \times \left[\frac{x^2+\left(\frac{\lambda^{\prime }}{2}\right)^2}{\bar{\epsilon}_{\lambda}(x,y)}-1 \right] \left(1-k_F\frac{\partial \ln(\beta(k_F))}{\partial k_F}\right), \!\!\!\!\!&& \label{b'_c3}
\end{flalign}
where we can make the change of variables $x = \lambda^{\prime} u$,  in $b^{\prime}_{c2}$ and $b^{\prime}_{c3}$. This allows us to prove that the coefficient $b^{\prime \prime}_{c2}$ scales as $\lambda'^2$, implying that this coefficient yields the following
\begin{equation}
b^{\prime}_{c2} = 0,
\label{b'_c2_final}    
\end{equation}
when we set $\lambda_c \to 0$ at the end of the calculation. For $b^{'}_{c3}$, however, we obtain the following expression
%
\begin{align}
b^{\prime}_{c3} =& -\frac{e^2 m^2}{\pi^4 k_F^2} \int^{\infty}_0 du \int^{\infty}_0 dy \frac{u C(\lambda^{\prime} u,y)}{[u^2+\frac{1}{4}]^2} \nonumber \\& \times\left[\frac{\lambda^{\prime 2}(u^2+\frac{1}{4})}{\lambda^{\prime 2}(u^2+\frac{1}{4})+\frac{\alpha r_s }{4 \pi}Q(\lambda^{\prime}u,y)}-1 \right] \nonumber \\ &
\times\left(1-k_F\frac{\partial \ln(\beta(k_F))}{\partial k_F}\right).
\label{b'_c3_2}    
\end{align}
%
At this point, we can perform a Taylor expansion in the $\lambda^{\prime} u \to 0$ limit, where the only term that will yield a non-zero contribution after setting $\lambda_c = 0$ at the end of the calculation comes from the expression
\begin{equation}
b^{'}_{c3} = \frac{e^2 m^2}{\pi^4 k_F^2 } \int^{\infty}_0 \!\!\!\!\!du \int^{\infty}_0 \!\!\!\!\!dy \frac{u C(0,y)}{[u^2+\frac{1}{4}]^2}\left(\!1\!-\!k_F\frac{\partial \ln(\beta(k_F))}{\partial k_F}\!\right),
\label{b'_c3_3}    
\end{equation}
since the remaining terms yield higher-order powers of $\lambda^{\prime}$. At this stage, it is straightforward to perform the integration over the $u$ variable, obtaining the following result
\begin{equation}
b^{\prime}_{c3} = \frac{5 e^2 m^2}{36 \pi^3 k_F^2}\left[1-k_F\frac{\partial \ln(\beta(k_F))}{\partial k_F}\right],
\label{b'_c3_final}
\end{equation}
where we have used
\begin{equation}
\int^{\infty}_0 dy C(0,y) = \frac{5 \pi}{72}.\label{integral_C(0,y)} 
\end{equation}

For $b^{\prime \prime}_c$, we separate the integral expression given by Eq.~\ref{b''_c_2d} as follows
\begin{align}
    b^{\prime \prime}_c &= b^{\prime \prime }_{c1} + b^{\prime \prime}_{c2},\label{b''_c_separation}\\
    b^{\prime \prime }_{c1} &= -\frac{2 e^2 m^2}{3 k_F^2 \pi^4} \int^{\infty}_0 dy \int^{\infty}_0 dx \frac{xI^{\prime}_2(x,y)}{\bar{\epsilon}_{\lambda}(x,y)},\label{b''_c1}\\
    b^{\prime \prime}_{c2} &= \frac{2 e^2 m^2}{3 k_F^2 \pi^4} \int^{\infty}_0dx\int^{\infty}_0 dy \frac{x I^{\prime}_2(x,y)}{x^2+\left(\frac{\lambda^{\prime}}{2} \right)^2},\label{b''_c2}
\end{align}
where we must focus first on the $b^{\prime \prime}_{c2}$ coefficient, as we cannot set $\lambda_c \to 0$ prior to performing the two-dimensional integration (otherwise, the integral would diverge). In this case, we exploit a property of $I^{\prime}_2(x,y)$ when integrating over the $y$ variable. We have
\begin{equation}
\int^{\infty}_0 dy I^{\prime}_2(x,y) = 0,
\label{integral_y_I'_2}
\end{equation}
for every $x < 1$, which implies that the integral expression for $b^{''}_{c2}$ directly reduces to an integration over Region 2. Thus we have
\begin{equation}
    b^{\prime \prime}_{c2} = \frac{2 e^2 m^2}{3 k_F^2 \pi^4}\int^{\infty}_0 dy \int^{\infty}_1 dx \frac{I_2^{\prime}(x,y)}{x},
    \label{b''_c2_final_integral}
\end{equation}
where, at this point, the above integral no longer requires a regulator to guarantee convergence, as this integration is restricted to Region $2$. 

\begin{figure*}[htp]
   \begin{center}
   \includegraphics[scale=0.7]{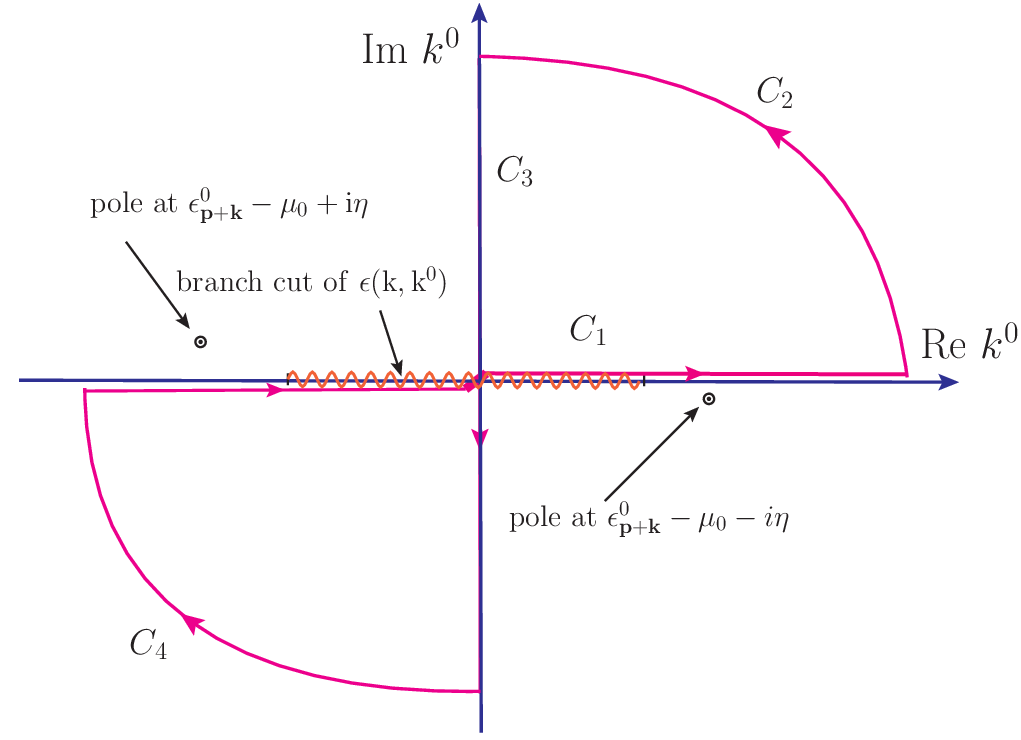} 
   \end{center}
   \caption{Complex frequency plane integration path for the ring-diagram series self-energy.}
   \label{self_ring_poles}
\end{figure*}

Now we can proceed to the second step of our systematic method, which allow us to set $\lambda_c =0$ before calculating the integrals defining $b^{\prime}_{c1}$, $b^{\prime \prime}_{c1}$, $b^{\prime \prime \prime}_{c}$, and $\Sigma_r(k_F,0)$. We start by discussing the integral expressions for the self-energy value $\Sigma_r(k_F,0)$ first.

Using the expression that defines $\Sigma_r(p,p^0)$ in Eq.~\ref{self-energy_ring}, setting the frequency variable $p^0=0$ allows us to map the integration over the real frequency variable $k^0$ to the imaginary frequency $i\nu$. This is possible by using Cauchy's theorem by performing a deformation of the complex contour path illustrated in Fig.~\ref{self_ring_poles}, which does not enclose any complex pole corresponding to the non-interacting Green's function $G^{0}(\vec{p}+\vec{k},k^0)$, as discussed in Appendix~\ref{Explicit_steps_Sigmar}. After this, we exploit a symmetry under sign inversion over the $x$ variable and subsequently rescale the variables in the following sequence: $\vec{k} \to k_F \vec{k}\,', \nu \to k_F^2\nu\,'/m$, then we perform the change of variable $\nu^{\prime} \to k^{\prime} y$ and finally set $k' \to 2x$. This allows us to obtain a compact expression where it is straightforward to calculate the integral over the azimuthal and polar angles, giving the following integral expression
\begin{align}
\Sigma_{r}(k_F,0) &= \Sigma_{r1}(k_F,0)+\Sigma_{r2}(k_F,0) \label{Sigma_r_decomposition},\\
\Sigma_{r1}(k_F,0) &= -\frac{e^2}{4 \pi^3 a_B} \int^{\infty}_0 \!\!\!dy \int^{\frac{1}{2}}_0 \!\!\!dx \frac{Q(x,y)g(x,y)}{\bar{\epsilon}(x,y)}, 
\label{self_r1_2d}\\
\Sigma_{r2}(k_F,0) &= -\frac{e^2}{4 \pi^3 a_B} \int^{\infty}_0 \!\!\!dy \int^{\infty}_{\frac{1}{2}} \!\!\!dx \frac{Q(x,y)g(x,y)}{\bar{\epsilon}(x,y)}, 
\label{self_r2_2d}
\end{align}
where $Q(x,y)$ [$g(x,y)$] is an even [odd] function in the $x$ variable. We can write the Taylor expansion of $g(x,y)$ as follows
\begin{equation}
g(x,y) \stackrel{x \to 0}{=} \sum^{\infty}_{n=0} \frac{x^{2n+1}}{(2n+1)!}g^{(2n+1,0)}(0,y), 
\label{g_taylor_x}    
\end{equation}
where $b^r_c$ is now separated as the following sum of two coefficients
\begin{align}
b^{r}_c &= b^r_{c,1} + b^{r}_{c,2},\label{b^r_c_sum_two}\\
b^r_{c,j} &= -\frac{\Sigma_{rj}(k_F,0)m^2}{12 \pi^2 k_F^3}, 
\end{align}
for $j=1,2$ and its associated function $f_3(x,y)$ is obtained from the product of the two functions in the numerator of Eq.~\ref{self_r1_2d}, yielding the following
\begin{equation}
f_3(x,y) = Q(x,y)g(x,y).
\label{f3}    
\end{equation}

The remaining $q^2$ coefficients, $b^{'}_{c1}$, $b^{''}_{c1}$, and $b^{\prime \prime \prime}_{c}$ are separated into Region 1 and Region 2 as follows
\begin{align}
b^{'}_{c1} &= b^{\prime}_{c11}+b^{\prime}_{c12},\label{b'_c1_regions}\\
b^{\prime \prime}_{c1} &= b^{\prime \prime}_{c11} + b^{\prime \prime}_{c12} + b^{\prime \prime}_{c2},\label{b''_c1_regions}\\
b^{\prime \prime \prime}_{c} &= b^{\prime \prime \prime}_{c,1}+b^{\prime \prime \prime}_{c,2},\label{b'''_xc_regions}
\end{align}
where the terms from Eqs.~\ref{b'_c1_regions}--\ref{b'''_xc_regions}, with the exception of $b^{\prime \prime}_{c2}$, are given by the following expressions
\begin{align}
b^{'}_{c11} =& -\frac{e^4m^3}{2 \pi^5 k^3_F} \int^{\infty}_0 dy \int^{\frac{1}{2}}_0 dx \frac{x C(x,y)Q(x,y)}{[\bar{\epsilon}_{\lambda}(x,y)]^2},\label{b'_c11}\\
b^{'}_{c12} =& -\frac{e^4m^3}{2 \pi^5 k^3_F} \int^{\infty}_0 dy \int^{\infty}_{\frac{1}{2}} dx \frac{x C(x,y)Q(x,y)}{[\bar{\epsilon}_{\lambda}(x,y)]^2},\label{b'_c12}\\
b^{\prime \prime }_{c11} =& -\frac{2 e^2 m^2}{3 k_F^2 \pi^4} \int^{\infty}_0 dy \int^{\frac{1}{2}}_0 dx \frac{xI^{\prime}_2(x,y)}{\bar{\epsilon}_{\lambda}(x,y)},\label{b''_c11}\\
b^{\prime \prime }_{c12} =& -\frac{2 e^2 m^2}{3 k_F^2 \pi^4} \int^{\infty}_0 dy \int^{\infty}_{\frac{1}{2}} dx \frac{xI^{\prime}_2(x,y)}{\bar{\epsilon}_{\lambda}(x,y)},\label{b''_c12}\\
b^{'''}_{c,1} =& \frac{m^3 e^4}{96 \pi^5 k_F^3}\int^{\frac{1}{2}}_0 \frac{dx}{x}\int_0^{\infty} dy \left(\frac{g(x,y)}{\overline{\epsilon_{\lambda}}(x,y)}\right)^2\nonumber \\
&\times  \left[1-\frac{\alpha r_s M_9(x,y)}{2 \pi \overline{\epsilon_{\lambda}}(x,y)}-\frac{(\alpha r_s M_{10}(x,y))^2}{8 \pi^2 (\overline{\epsilon_{\lambda}}(x,y))^2} \right],\label{b'''_xc_2d_region1}\\
b^{'''}_{c,2} =& \frac{m^3 e^4}{96 \pi^5 k_F^3}\int^{\infty}_{\frac{1}{2}} \frac{dx}{x}\int_0^{\infty} dy \left(\frac{g(x,y)}{\overline{\epsilon_{\lambda}}(x,y)}\right)^2\nonumber \\
&\times  \left[1-\frac{\alpha r_s M_9(x,y)}{2 \pi \overline{\epsilon_{\lambda}}(x,y)}-\frac{(\alpha r_s M_{10}(x,y))^2}{8 \pi^2 (\overline{\epsilon_{\lambda}}(x,y))^2} \right],\label{b'''_xc_2d_region2}
\end{align}
where $f_1(x,y)$ and $f_2(x,y)$ are straightforwardly extracted from Eqs.~\ref{b'_c11} and~\ref{b''_c11}, yielding the following expressions
\begin{align}
    f_1(x,y) &= x C(x,y) Q(x,y),
    \label{f1}\\
    f_2(x,y) &= xI^{\prime}_2(x,y),
    \label{f2}
\end{align}
where the functions associated with the integral expressions of $b^{\prime \prime \prime}_{c,1}$ and $b^{\prime \prime \prime}_{c,2}$ from Eqs.~\ref{b'''_xc_2d_region1} and~\ref{b'''_xc_2d_region2}, respectively, are less immediate to determine, as an algebraic manipulation is required to express these coefficient in terms of the integral kinds $L^{1}_{im}(r_s)$ and $L^{2}_{im}(r_s)$ in the subsequent step of our systematic approach.

Now we apply step 3, where $b^{\prime}_{c,1j}$, $b^{\prime \prime}_{c,1j}$, $b^{\prime \prime \prime}_{c,j}$ and $b^r_c$ (for $j=1,2$) are expressed in terms of the integral kinds $L^{j}_{im}(r_s)$. These contributions to $b_c$ are given by the following expressions
\begin{align}
b^{'}_{c,1j} &= -\frac{e^4 m^3}{2 \pi^5 k_F^3}L^j_{12}(r_s),\label{b'_c11_Lform}\\
b^{''}_{c,1j} &= -\frac{2 e^2 m^2}{3\pi^4k_F^2}L^j_{21}(r_s),\label{b''_c11_Lform}\\
b^r_{c,j} &= \frac{m^3 e^4}{48 \pi^5 k_F^3} L^j_{31}(r_s),\label{b^r_c_Lform}\\
b^{\prime \prime \prime}_{c,j} &= \sum^{6}_{n=1} b^{\prime \prime \prime}_{c,jn},\label{b'''_xc_sum}
\end{align}
where $b^{\prime \prime \prime}_{c,jn}$ ($n=1,\dots,6$) are given in terms of the integral kinds $L^j_{nm}(r_s)$ as follows
\begin{align}
b^{\prime \prime \prime}_{c,j1} &= -\frac{m^3 e^4}{192 \pi^5 k_F^3}L^j_{42}(r_s),\label{b'''_xc_11_Lform}\\
b^{\prime \prime \prime}_{c,j2} &= \frac{m^3 e^4}{16 \pi^5 k_F^3}L^j_{53}(r_s),\label{b'''_xc_12_Lform}\\
b^{\prime \prime \prime}_{c,j3} &= -\frac{3m^3 e^4}{64 \pi^5 k_F^3}L^j_{64}(r_s),\label{b'''_xc_13_Lform}
\end{align}
\begin{align}
b^{\prime \prime \prime}_{c,j4} &= \frac{m^3 e^4 \alpha r_s}{96 \pi^6 k_F^3} L^{j}_{73}(r_s),\label{b'''_xc_14_Lform}\\
b^{\prime \prime \prime}_{c,j5} &= -\frac{m^3 e^4 \alpha r_s}{64 \pi^6 k_F^3}L^j_{84}(r_s),\label{b'''_xc_15_Lform}\\
b^{\prime \prime \prime}_{c,j6} &= -\frac{m^3 e^4 (\alpha r_s)^2}{768 \pi^7 k_F^3}L^j_{94}(r_s),\label{b'''_xc_16_Lform}
\end{align}
where the $L^1_{nm}(r_s)$ integral kinds are evaluated by using Eq.~\ref{L1_im_l=0}. The functions $f_n(x,y)$ associated with these $L^1_{nm}(r_s)$ integral kinds are given by the following expressions
\begin{align}
f_4(x,y) =& \frac{[g(x,y)]^2}{x}, \label{f4}\\
f_5(x,y) =& x[g(x,y)]^2,\label{f5}\\
f_6(x,y) =& x^3[g(x,y)]^2,\label{f6}\\
f_7(x,y) =& f_4(x,y)h_1(x,y),\label{f7}\\
h_1(x,y) =& 1 + \frac{(y^2-3x^2)}{8x}g(x,y)\nonumber \\
&-\frac{(y^2+x^2)(3y^2+1-x^2)}{2D(x,y)},\label{h1}\\
f_8(x,y) =& f_5(x,y)h_2(x,y),\label{f8}\\
h_2(x,y) =& 2-\frac{(y^2+x^2)}{2x}g(x,y),\label{h2}\\
f_9(x,y) =& f_4(x,y)h_3(x,y),\label{f9}\\
h_3(x,y) =& \left[h_2(x,y) \right]^2,\label{h3}
\end{align}
where some of the terms of the Taylor expansion for each $f_n(x,y)$ function used in this work are provided in Appendix~\ref{list_functions_used}.

The final step is to substitute the $r_s$-expanded forms of the integral kinds $L^1_{nm}(r_s)$ and $L^2_{nm}(r_s)$. The NLO terms in $r_s$ arise from $b^{\prime}_{c11}$, $b^{\prime \prime}_{c11}$, and $b^{\prime \prime \prime}_{c1}$. In this work, we focus exclusively on the NLO $r_s$ contribution, which in this case turns out to be $r_s^3 \ln(r_s)$, stemming from the integral kinds $L^1_{nm}(r_s)$. This can be seen by observing that the prefactor of each coefficient expression for $b^{\prime}_{c,12}, b^{r}_{c,2}$, and $b^{\prime \prime \prime}_{c,2}$ corresponding to Region 2 (given in Eq.~\ref{b'_c11_Lform} and Eqs.~\ref{b^r_c_Lform}--\ref{b'''_xc_16_Lform}) contributes as $r_s^3$, while the integral kind $L^2_{i,m}(r_s)$ yields a constant as its leading order in $r_s$. Additionally, the combination of $b^{\prime \prime}_{c12}$ and $b^{\prime \prime}_{c2}$ yields the following expression
\begin{align}
b^{\prime \prime}_{c,122} &= b^{\prime \prime}_{c12} + b^{\prime \prime}_{c2},\label{b''_c_122_sum}\\ 
b^{\prime \prime}_{c,122} &= \frac{e^4 m^3}{6 k_F^3 \pi^5}\int^{\infty}_0 dy \int^{\infty}_{\frac{1}{2}} dx \frac{I^{\prime}_2(x,y)Q(x,y)}{x\bar{\epsilon}(x,y)},\label{b''_c_122_integral}
\end{align}
where we can perform a Taylor expansion of $1/\bar{\epsilon}(x,y)$ in the $r_s \to 0$ limit, analogous to our treatment of the integral kind $L^2_{im}(r_s)$, which yields a constant as its leading-order contribution in $r_s$. This demonstrates that the coefficient $b^{\prime \prime}_{c,122}$ from Eq.~\ref{b''_c_122_integral} contributes terms at order $r_s^3$ and higher.
Through this procedure, we obtain the following results
\begin{align}
b^{\prime}_{c} &\approx  b^{\prime (\mathrm{LO})}_{c} + C^{\prime}_{L3} r^3_s \ln(r_s) + O(r_s^3),
    \label{b'_c_final_NLO_result}\\
b^{\prime \prime}_{c} &\approx b^{\prime \prime (\mathrm{LO})}_c + C^{\prime \prime}_{L3} r_s^3\ln(r_s) + O(r_s^3),\label{b''_c_NLO_final_result}\\
b^{\prime \prime \prime}_{c} &\approx b^{\prime \prime \prime (\mathrm{LO})}_{c} + C^{\prime \prime \prime}_{L3} r_s^3 \ln(r_s) + O(r_s^3),\label{b'''_c_NLO_final_result}\\
b^r_c &\approx C^{r}_{L3} r^3_s \ln(r_s) + O(r_s^3), \label{b_r_NLO_final_result}
\end{align}
where $b^{\prime (\mathrm{LO})}_c$, $b^{\prime \prime (\mathrm{LO})}_{c}$ and $b^{\prime \prime \prime (\mathrm{LO})}_{c}$ are given by Eqs.~\ref{b'_c_final_result},~\ref{b''_c_final_result} and~\ref{b'''_c_final_result}, respectively, and the coefficients $C^{\prime}_{L3}$, $C^{\prime \prime}_{L3}$, $C^{\prime \prime \prime}_{L3}$, and $C^r_{L1}$ are obtained by calculating the following integrals
%
\begin{align}
C^{\prime}_{L3} =& \frac{e^4m^3 (\alpha a_B)^3}{8 \pi^5} \int^{\infty}_0 \!\!\!dy \left[C^{(2,0)}(0,y)Q(0,y)\right. \nonumber \\
& \left. +\;C(0,y)Q^{(2,0)}(0,y) \right],
\label{C'_L3_integral_expression}\\
C^{\prime \prime}_{L3} =& -\frac{e^4m^3 (\alpha a_B)^3}{24 \pi^5} \int^{\infty}_0 \!\!\!dy \left[Q^{(2,0)}(0,y) I^{\prime}_2(0,y)\right. \nonumber \\
 & \left.+\;Q(0,y)I^{\prime (2,0)}_2(0,y) \right],\label{C''_L3_integral_expression}\\
C^{r}_{L3} =& -\frac{e^4m^3 (\alpha a_B)^3}{24 \pi^5} \int^{\infty}_0 dy \frac{Q(0,y)}{1+y^2}, \label{Cr_L3_integral_expression}
\end{align}
%
where we identify that the coefficient $C^{r}_{L3}$ arises solely from the $\mathcal{D}$-type contributions, whereas $C^{\prime}_{L3}$ and $C^{\prime \prime}_{L3}$ originate from both $\mathcal{D}$-type and $\mathcal{H}$-type terms. The term $C^{\prime \prime \prime}_{L3}$ is given by the sum of the following integrals
\begin{align}
C^{\prime \prime \prime}_{L3} &= \sum^3_{i=1} C^{\prime \prime \prime}_{L3,i},
\label{C'''_L3_sum_integrals}\\
C^{\prime \prime \prime}_{L3,1} &= -\frac{e^4m^3 (\alpha a_B)^3}{36 \pi^5} \int^{\infty}_0 dy \frac{3y^2-1}{(1+y^2)^4},
\label{C'''_L31_integral_expression}\\
C^{\prime \prime \prime}_{L3,2} &= \frac{e^4m^3 (\alpha a_B)^3}{3 \pi^5} \int^{\infty}_0 dy \frac{3y^2-1}{(1+y^2)^4},
\label{C'''_L32_integral_expression}\\
C^{\prime \prime \prime}_{L3,3} &= -\frac{e^4m^3 (\alpha a_B)^3}{4 \pi^5} \int^{\infty}_0 dy \frac{3y^2-1}{(1+y^2)^4},
\label{C'''_L33_integral_expression}
\end{align}
where the coefficients $C^{\prime \prime \prime}_{L3,j}$ come solely from the terms $b^{\prime \prime \prime}_{xc,1j}$ (for $j=1,2,3$) and represent $\mathcal{D}$-type contributions to $b^{\prime \prime\prime}_{c}$. The remaining coefficients $b^{\prime \prime \prime}_{c,1j}$ (for $j=4,5,6$) contribute only to higher-order powers of $r_s$, which can be extracted from both $\mathcal{H}$ and $\mathcal{D}$-type terms in $r_s$.

We now explain in more detail the calculation of the integral kinds $L^1_{im}(r_s)$ used for the calculation of the coefficients $b^{\prime}_{c1}$, $b^{\prime \prime}_{c,1}$, $b^{r}_{c1}$ and $b^{\prime \prime \prime}_{c,1}$ in their respective subsections.

\subsection{Calculation for $b'_{c,1}$}
\label{calculation_for_b'_c1}
To calculate the coefficient of the $r_s^3 \ln(r_s)$ term for $b^{\prime}_{c,1}$, we use Eq.~\ref{L1_im_l=0} up to $n=1$ in the sum when substituting it into the expression of $b^{\prime}_{c,11}$ given by Eq.~\ref{b'_c11_Lform} with $j=1$. In Region 1 of integration, we have
%
\begin{align}
b^{\prime}_{c,11} \approx& -\frac{e^4m^3}{2 \pi^5 k_F^3}\int^{\infty}_0 \!\!\!dy\left[\Gamma^1_{02}(r_s,y)f^{(1,0)}_1(0,y) \right. \nonumber \\ 
& \left. +\;\frac{\Gamma^{3}_{02}(r_s,y)f^{(3,0)}(0,y)}{3!}+... \right].
\label{b'_c1_sum_terms}    
\end{align}
%
By using the expressions for $f^{(1,0)}(0,y)$ and $f^{(3,0)}(0,y)$ given in Appendix~\ref{list_functions_used}, together with the integral kinds $\Gamma^1_{02}(r_s,y)$ and $\Gamma^3_{02}(r_s,y)$ given in Appendix~\ref{Expressions_Gamma_used}, we obtain the following expression for $b^{\prime}_{c,11}$
\begin{equation}
b^{\prime}_{c,11} \approx -b^{\prime (\mathrm{LO})}_c+C^{\prime}_{L3} r_s^3 \ln(r_s)+...,
\label{b'_c11_expansion_rs}    
\end{equation}
where $b^{\prime (\mathrm{LO})}_c$ is the coefficient previously obtained by Ma-Brueckner in Ref.~\onlinecite{Ma-Brueckner}. This value and the coefficient $C^{\prime}_{L3}$ are obtained from the following integrals
\begin{align}
b^{\prime (\mathrm{LO})}_{c} &= \frac{e^2 m^2}{\pi^4 k_F^2} \int^{\infty}_0 dy C(0,y) = \frac{5 e^2 m^2}{72 \pi^3 k_F^2},
    \label{b'_c_MB}\\
C^{\prime}_{L3} &= \frac{e^4m^3 (\alpha a_B)^3}{8 \pi^5} \left[\mathcal{V}_1+\mathcal{V}_2 \right],
\label{C'_L3_integral_expression}    
\end{align}
where $\mathcal{V}_1$ and $\mathcal{V}_2$ are given by the following integrals
\begin{align}
 \mathcal{V}_1 &= \int^{\infty}_0 dy C^{(2,0)}(0,y)Q(0,y),\label{V1}\\
 \mathcal{V}_2 &= \int^{\infty}_0 dy C(0,y)Q^{(2,0)}(0,y).\label{V2}
\end{align}
The integral $\mathcal{V}_1$ is obtained from the sum of five integrals given by
\begin{align}
\mathcal{V}_1 &= \frac{8}{3}\mathcal{D}_{8,5} + 2\mathcal{D}_{6,5}-\frac{70}{9}\mathcal{D}_{4,5}-2\mathcal{D}_{2,5}-2\mathcal{D}_{0,5} \nonumber \\
&= \frac{2 \pi}{9}\left[\frac{625}{192}-6\ln(2)\right],
\label{V1_sum_expression}    
\end{align}
where the integral kinds $\mathcal{D}_{n,m}$ are defined by the following integral expression
\begin{align}
\mathcal{D}_{n,m} &\equiv \int^{\infty}_0 dy \frac{y^n R(y)}{(1+y^2)^m},
\label{D_nm_integral}\\
R(y) &= 1-y\tan^{-1}\left(\frac{1}{y}\right).\label{R}
\end{align}

The integrals $\mathcal{D}_{nm}$ are calculated explicitly in Appendix~\ref{steps_Dnm_beta_n}, and their resulting values are summarized in Table~\ref{table_Dnm}.
\def\arraystretch{2}
\begin{table}[ht]
    \centering
    \begin{tabular}{|c|c|c|c|c|}
        \hline
          $\mathcal{D}_{0,5}$ & $\mathcal{D}_{2,5}$  & $\mathcal{D}_{4,5}$ & $\mathcal{D}_{6,5}$ & $\mathcal{D}_{8,5}$ \\
        \hline
         $\frac{187\pi}{2048}$ & $\frac{47\pi}{6144}$ & $\frac{17 \pi}{6144}$ & $\frac{5\pi}{2048}$ & $\pi\left[\frac{2161}{6144}-\frac{\ln(2)}{2}\right]$ \\
        \hline
    \end{tabular}
    \caption{Values of the integrals $\mathcal{D}_{n,m}$ obtained by using Eq.~\ref{D_nm_integral}.}
    \label{table_Dnm}
\end{table}

The integral $\mathcal{V}_2$ can be written in terms of the integral kinds $\beta_n$
\begin{equation}
\mathcal{V}_2 = -\frac{2}{3}\beta_3+\frac{10}{27} \beta_4+\frac{8}{27}\beta_5 = -\frac{23 \pi}{864},
    \label{V2}
\end{equation}
where the integral kinds $\beta_n$ are defined by
\begin{equation}
\beta_n = \int^{\infty}_0 dy \frac{1}{(x^2+1)^n},
\label{beta_n}    
\end{equation}
where the values of $\beta_n$ used to calculate $\mathcal{V}_2$ are listed in Table~\ref{beta_n}. The detailed calculation of these integral kinds is given in Appendix~\ref{steps_Dnm_beta_n}.

\def\arraystretch{1.5}
\begin{table}[ht]
    \centering
    \begin{tabular}{|c|c|c|c|c|}
        \hline
         $\beta_1$& $\beta_2$ &$\beta_{3}$ & $\beta_4$  & $\beta_5$ \\
        \hline
         $\pi/2$& $\pi/4$ &$3\pi/16$ & $5\pi/32$ & $35\pi/256$  \\
        \hline
    \end{tabular}
    \caption{Values of the integrals $\beta_{n}$ obtained by using Eq.~\ref{beta_n}.}
    \label{table_beta_n}
\end{table} 

Using the values of $\mathcal{V}_1$ and $\mathcal{V}_2$, we obtain the coefficient $C^{\prime}_{L3}$
\begin{equation}
C^{\prime}_{L3} = \frac{e^4m^3 (\alpha a_B)^3}{\pi^4} \left[\frac{301}{3456}-\frac{\ln(2)}{6} \right].
\label{C'_L3_value}    
\end{equation}

\subsection{Calculation for $b^{\prime \prime}_{c,1}$}
\label{Calculation_for_b''_c}
For the calculation of the coefficient of the $r_s^3 \ln(r_s)$ term for $b^{\prime \prime}_{c,1}$, we use Eq.~\ref{L1_im_l=0}, retaining terms up to $n=1$ in the summation when substituting it into Eq.~\ref{b''_c11_Lform} for $j=1$. This yields
%
\begin{align}
b^{\prime \prime}_{c,11} \approx& -\frac{2 e^2 m^2}{3 \pi^4 k_F^2}\int^{\infty}_0 \!\!\!dy\left[\Gamma^1_{01}(r_s,y)f^{(1,0)}_2(0,y)\right. \nonumber \\
&\left.+\;\frac{\Gamma^{3}_{01}(r_s,y)f^{(3,0)}_2(0,y)}{3!}+... \right],
\label{b''_c1_sum_terms}    
\end{align}
%
where this integral reduces to the sum of the following terms
\begin{align}
    b^{\prime \prime}_{c,1} &= b^{\prime \prime (\mathrm{LO})}_c+C^{\prime \prime}_{L3} r_s^3\ln(r_s)+...,
    \label{b''_c1_sum_terms}\\
    C^{\prime \prime}_{L_3} &= -\frac{e^4m^3 (\alpha a_B)^3}{24 \pi^5}\left[\mathcal{V}_3+\mathcal{V}_4 \right],\\
    \mathcal{V}_3 &\equiv \int^{\infty}_0 dy Q^{(2,0)}(0,y) I^{\prime}_2(0,y),\label{V3}\\
    \mathcal{V}_4 &\equiv \int^{\infty}_0 dy Q(0,y)I^{\prime (2,0)}_2(0,y),\label{V4}
\end{align}
where the first integral expression for $b^{\prime \prime (\mathrm{LO})}_c$ yields the Ma-Brueckner value reported in Ref.~\onlinecite{Ma-Brueckner},  given by
%
\begin{align}
    b^{\prime \prime (\mathrm{LO})}_c &= \frac{e^2m^2}{3 \pi^4 k_F^2}\int^{\infty}_0 \!\!\!dy \left[I^{\prime}_2(0,y)\ln\left|Q(0,y) \right|-\frac{I^{\prime (2,0)}_2(0,y)}{8} \right] \nonumber \\
    &\approx \frac{e^2 m^2}{\pi^3k_F^2}(0.10359),
    \label{b''_c_MB}
\end{align}
%
where we have exploited the following property of the function $I^{\prime}_2(0,y)$
\begin{equation}
\int^{\infty}_0 dy I^{\prime (2,0)}_2(0,y) = 0.
    \label{I'_2_integral_(1,0)}
\end{equation}

The integral kinds $\mathcal{V}_3$ and $\mathcal{V}_4$ can be expressed in terms of the integral kinds $\beta_n$ as follows
\begin{eqnarray}
\mathcal{V}_3 &=& 2\beta_3-10\beta_4+8\beta_5 = -\frac{3 \pi}{32},
   \label{V3_sum_expression}\\
   \mathcal{V}_4 &=& -6\mathcal{D}_{0,5}+30\mathcal{D}_{2,5}-110\mathcal{D}_{4,5}+50\mathcal{D}_{6,5}+4\mathcal{D}_{8,5}\nonumber \\
   &=& \pi\left[\frac{29}{32}-2\ln(2) \right].\label{V4_sum_expression}
\end{eqnarray}
Combining the results for $\mathcal{V}_3$ and $\mathcal{V}_4$, we obtain the following value for the coefficient $C^{\prime \prime}_{L3}$
\begin{equation}
C^{\prime \prime}_{L3} = \frac{e^4m^3(\alpha a_B)^3}{384 \pi^4} \left[32\ln(2)-13 \right].
    \label{C''_L3_value}
\end{equation}

\subsection{Calculation for $b^r_{c,1}$}
For the calculation of the coefficient of the $r_s^3\ln(r_s)$ term in $b^{r}_c$, we use Eq.~\ref{L1_im_l=0}, retaining only the $n=0$ term in the summation when substituting it into Eq.~\ref{b^r_c_Lform} for $j=1$. We have
\begin{equation}
b^{r}_{c,1} = \frac{e^4m^3}{48\pi^5k_F^3}\int^{\infty}_0 dy\left[\Gamma^1_{01}(r_s,y)f^{(1,0)}_3(0,y)+... \right],
    \label{b^r_c_sum_terms}
\end{equation}
where this integral reduces to the following expression in terms of the integral kind $\mathcal{V}_5$. We have
\begin{eqnarray}
b^{r}_{c,1} &=& C^{r}_{L3} r_s^3\ln(r_s)+...,\label{b^r_c_rs} \\
C^r_{L3} &=& -\frac{e^4m^3 (\alpha a_B)^3}{6 \pi^5}\mathcal{D}_{0,1},\label{C^r_L3_D01}
\end{eqnarray}
where the integral kind $\mathcal{D}_{0,1}$ is given by the following value
\begin{equation}
\mathcal{D}_{0,1} = \frac{\pi}{2}\left[1-\ln(2) \right].
\label{D01}    
\end{equation}
By using this value, we find that the $C^{r}_{L3}$ coefficient is given by
\begin{equation}
C^{r}_{L3} = \frac{e^4 m^3 (\alpha a_B)^3}{12 \pi^4} \left[\ln(2)-1 \right].
\label{Cr_L3_value}    
\end{equation}

\subsection{Calculation for $b^{\prime \prime \prime}_{c,1}$}
\label{Calculation_for_b'''_xc1}
 For the calculation of $b^{\prime \prime \prime}_{c,1}$ by using Eqs.~\ref{b'''_xc_sum}-~\ref{b'''_xc_16_Lform}, where we can notice directly that by doing a power counting of the $r_s$ terms in the prefactors of the integral expressions of $b^{\prime \prime \prime}_{c,1j}$, only the particular case ($j=1,2,3$) yields the $r_s^3\ln(r_s)$ term. The integral expressions that we have for these three terms contributing to $b^{\prime \prime \prime}_{c,1}$ are given by 
\begin{align}
  b^{\prime \prime \prime}_{c,11} &= -{{\theta} \over 3} \int^{\infty}_0 \!\!\!\!\!\!dy
    \sum^1_{n=0}\frac{\Gamma^{2n+1}_{02}(r_s,y)f^{(2n+1,0)}_4(0,y)}{(2n+1)!}, \label{b'''_xc,11_integral_form}\\
b^{\prime \prime \prime}_{c,12} &= 4 \theta \int^{\infty}_0  \!\!\!dy\sum^2_{n=0}\frac{\Gamma^{2n+1}_{03}(r_s,y)f^{(2n+1,0)}_5(0,y)}{(2n+1)!},\label{b'''_xc,12_integral_form}\\
b^{\prime \prime \prime}_{c,13} &= -{3 \theta} \int^{\infty}_0  \!\!\! dy\sum^{3}_{n=0}\frac{\Gamma^{2n+1}_{04}(r_s,y)f^{(2n+1)}_6(0,y)}{(2n+1)!},\label{b'''_xc,13_integral_form}
\end{align}
where $\theta = \frac{m^3 e^4}{64 \pi^5 k_F^3}$ and we have ignored higher-order
terms. We have listed the expressions of the integral kinds $\Gamma^n_{im}(r_s,y)$ in Appendix~\ref{Expressions_Gamma_used} and the terms $f^{(2n+1,0)}_i(0,y)$ in Appendix~\ref{list_functions_used}. For $b^{\prime \prime \prime}_{c,1j}$ (for $j=1,2,3$), we obtain the following $r_s$ expansion
\begin{equation}
b^{\prime \prime \prime}_{c,1j} = b^{\prime \prime \prime (\mathrm{LO})}_{c,j}+C^{\prime \prime \prime \prime}_{L3,j} r_s^3 \ln(r_s),\quad \mathrm{for\;}j=1,2,3,
    \label{b'''_xc_11_rs}
\end{equation}
where the terms $b^{\prime \prime \prime (\mathrm{LO})}_{c,j}$ contribute to the $b^{\prime \prime \prime}_{c}$ coefficient obtained by Ma-Brueckner in Ref.~\onlinecite{Ma-Brueckner}, and are given by
\begin{align}
b^{\prime \prime \prime (\mathrm{LO})}_{c,1} &=  -\frac{m^2e^2}{6\pi^4k_F^2}\int^{\infty}_0 dy \frac{1}{(1+y^2)^2Q(0,y)},
\label{b'''_xc_MB,1}\\
b^{\prime \prime \prime (\mathrm{LO})}_{c,2} &= \frac{m^2 e^2}{\pi^4k_F^2}\int^{\infty}_0 dy \frac{1}{(1+y^2)^2Q(0,y)},\label{b'''_xc_MB,2}\\
b^{\prime \prime \prime (\mathrm{LO})}_{c,3} &= -\frac{m^2 e^2}{2\pi^4k_F^2}\int^{\infty}_0 dy \frac{1}{(1+y^2)^2 Q(0,y)}.\label{b'''_xc_MB,3}
\end{align}
The coefficients $C^{\prime \prime \prime}_{L3,j}$ (for $j=1,2,3$) are obtained from the integral kind $\beta_n$
\begin{align}
C^{\prime \prime \prime}_{L3,1} &= -\frac{m^3 e^4 (\alpha a_B)^3}{36 \pi^5}\left[3\beta_3 -4\beta_4\right], 
\label{C'''_L3_1}\\
C^{\prime \prime \prime}_{L3,2} &= \frac{m^3 e^4 (\alpha a_B)^3}{3 \pi^5}\left[3\beta_3 -4\beta_4\right], 
\label{C'''_L3_2}\\
C^{\prime \prime \prime}_{L3,3} &= -\frac{m^3 e^4 (\alpha a_B)^3}{4 \pi^5}\left[3\beta_3 -4\beta_4\right].
\label{C'''_L3_3}\\
\end{align}

The rest of the terms $b^{\prime \prime \prime}_{c,1j}$ (for $j=4,5,6$) only contribute to the $r_s^2$ coefficient and higher-order terms than $r_s^3\ln(r_s)$.  We had
\begin{align}
b^{\prime \prime \prime (\mathrm{LO})}_{c,4} &= \frac{2}{3}\frac{m^2 e^2}{\pi^4 k_F^2}\int^{\infty}_0 dy \frac{2y^2+1}{(1+y^2)^4 Q^2(0,y)},
\label{b'''_xc_MB,4}\\
b^{\prime \prime \prime (\mathrm{LO})}_{c,5} &= -\frac{2}{3}\frac{m^2 e^2}{\pi^4 k_F^2} \int^{\infty}_0 dy \frac{1}{(1+y^2)^3 Q^2(0,y)},\label{b'''_xc_MB_5}\\
b^{\prime \prime \prime (\mathrm{LO})}_{c,6} &= -\frac{8m^2e^2}{9 \pi^4 k_F^2} \int^{\infty}_0 dy \frac{1}{(1+y^2)^4 Q^3(0,y)}.\label{b'''_xc_MB,6}
\end{align}

By combining all the contributions to $b^{\prime \prime \prime}_{c,1}$, we obtain the following final result
\begin{align}
b^{\prime \prime \prime}_{c,1} &= b^{\prime \prime \prime (\mathrm{LO})}_{c} + C^{\prime \prime \prime}_{L3}r_s^3\ln(r_s),
    \label{b'''_xc_final_rs}\\
    b^{\prime \prime \prime (\mathrm{LO})}_{c} &= 0.59136\frac{m^2 e^2}{(2\pi)^3k_F^2},\label{b'''_xc_final_rs2_}\\
    C^{\prime \prime \prime}_{L3} &= -\frac{m^3 e^4 (\alpha a_B)^3}{288 \pi^4},\label{C'''_L3_final}
\end{align}
where the value of $b^{\prime \prime \prime (\mathrm{LO})}_{c}$ was found by summing the integral expressions given by Eqs.~\ref{b'''_xc_MB,1}--\ref{b'''_xc_MB,3} and Eqs.~\ref{b'''_xc_MB,4}--\ref{b'''_xc_MB,6}, which allows the resulting expression to be re-expressed in a more compact form. This integral expression is given by
\begin{align}
b^{\prime \prime \prime (\mathrm{LO})}_{c} &= \sum^6_{j=1} b^{\prime \prime \prime (\mathrm{LO})}_{c,j}, 
    \label{b'''_xc_MB_sum_terms}\\
b^{\prime \prime \prime (\mathrm{LO})}_{c} &= \frac{m^2 e^2}{9\pi^4 k_F^2}\int^{\infty}_0 dy \frac{3+2y^2}{(1+y^2)^3 Q(0,y)},\label{b'''_xc_MB_compact_final} 
\end{align}
where this integral expression was obtained by exploiting the result of the following integral
\begin{equation}
  \int^{\infty}_0 \!\!\!\!\frac{dy}{(1+y^2)^4}\left[\frac{2y^2(1+y^2)}{3R(y)}\!+\!\frac{y^2}{ R^2(y)}\!-\!\frac{1}{3R^3(y)} \right] =0.
\label{b'''_xc_integral_part_used}    
\end{equation}
After combining all the expressions, we find that the values of   $b^{\prime (\mathrm{LO})}_c$, $b^{\prime \prime (\mathrm{LO})}_c$, $b^{\prime \prime \prime (\mathrm{LO})}_{c}$, and $b^r_c$ are those reported
in Table~\ref{table_b_c_MB}. Meanwhile, the coefficients of the $r_s^3 \ln(r_s)$ term (Eqs.~\ref{b'_c_final_NLO_result},\ref{b''_c_NLO_final_result},\ref{b'''_c_NLO_final_result}, and \ref{b_r_NLO_final_result}) are reported in Table~\ref{table_coeff_NLO}.

\subsection{Our calculation of the coefficient of the NLO term in $r_s$  is exact}
\label{LO_rs_higher_order_pixc}
We wish to show that no other proper-polarization diagrams contribute to the coefficient of $r_s^{3}\ln r_s$, which constitutes the NLO term in the $r_s$ expansion of $b_{xc}$. Within the many-body framework and RPA renormalization scheme discussed in Ref.~\onlinecite{benites2024}, we reorganize the terms arising in standard perturbation theory
(i.e., by re-summing the family of
diagrams which results in diagrams containing the
RPA-screened interaction $\tilde{V}_e(k^{\mu})$) when calculating the ground-state density-response function. Consequently, the resulting series of
diagrams contributing to the proper polarization function contains only
RPA-screened interaction lines.  In this RPA-based reorganized perturbative expansion (RPARPE), the $m$-th order is determined by the number of RPA-screened interaction lines in a diagram. We then contrast this diagrammatic hierarchy of the RPARPE with the resulting order in the $r_s$ expansion for each $q^2$ coefficient
arising from the $m$-th order proper-polarization function.  

We recall that the proper-polarization functions illustrated in Figs.~\ref{GA_Fock} and~\ref{Pi_c} were obtained within the RPARPE. As explained in section~\ref{integrals_b_xc}, both $\Pi^{xc}_1(q,0)$ and $\Pi^{xc}_2(q,0)$ correspond to specific combinations of the proper-polarization diagrammatic
components shown in Figs.~\ref{GA_Fock} and~\ref{Pi_c}.
Specifically, $\Pi^{xc}_1(q,0)$ and $\Pi^{xc}_2(q,0)$  constitute first-order contributions in the RPARPE, whereas $\Pi^{xc}_3(q,0)$ is a second-order contribution. In general, an $m$-th order diagram in the RPARPE captures terms of equal-or-higher-order in $r_s$ than its corresponding counterpart in the standard perturbation expansion. We aim to apply the RPARPE to systematically investigate any other equal-or-higher-order (EHO)
diagrammatic contributions to the proper-polarization function beyond those contained in $\Pi^{xc}_3(q,0)$; in addition, we wish to keep track of the
terms in their respective $r_s$ expansions in the high-density limit ($r_s \to 0$). We denote the sum of all such other EHO
contributions to the static proper-polarization function as $\Pi^{xc}_{EH}(q,0)$. Predicting even the LO $r_s$ contribution to $b_{xc}$ from an arbitrary $m$-th order proper-polarization diagram is nontrivial because simple dimensional analysis via
rescaling the integration variables is insufficient.
Instead, a rigorous analysis of the underlying integral expressions for each term in $\Pi^{xc}_{EH}(q,0)$ is required, mirroring the approach used in this work to extract the NLO $r_s$ contribution to $b_c$ by examining the integrand structures in Region $1$ via Eq.~\ref{b_c_def}.

\begin{figure*}[htp]
   \begin{center}
   \includegraphics[scale=0.6]{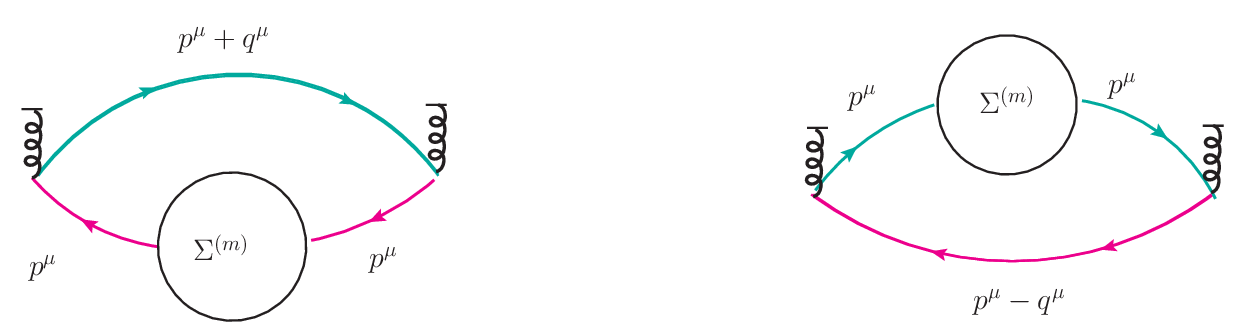} 
   \end{center}
   \caption{Family of diagrams corresponding to the irreducible polarization function of the $m$-th order of the RPA-based reorganized perturbative expansion. These $m$-th order diagrams contain one insertion of an $m$-th order self-energy at one of the two fermion propagators.}
   \label{family_polarization_selfenergy}
\end{figure*}

However, we can always find a set of diagrams that share the same LO contribution in $r_s$ in the high-density limit, even if they do not necessarily
belong to the same order within the RPARPE. In this work, we denote
such a set of diagrams as $\mathcal{S}_i$, where a higher index $i \in \mathbb{N}$ corresponds to a higher-order LO $r_s$ contribution deriving from
the proper-polarization diagrams within that specific set. In our notation, the set of diagrams contributing to $\Pi^{xc}_j(q,0)$ (for $j=1,2,3$) is labeled
as $\mathcal{S}_1$. Within every set $\mathcal{S}_i$, there is a representative diagram belonging to a family of proper-polarization diagrams with a distinct diagrammatic topology that is straightforward to track across RPARPE orders. The topology of this family of diagrams is readily conceived as the insertion of the $m$-th order self-energy into one of the two fermionic propagators inside a fermion loop, which arises upon applying Wick's theorem to the expression defining the density-response function. This family of diagrams is illustrated in Fig.~\ref{family_polarization_selfenergy}. Crucially, tracking the RPARPE order of this family of diagrams is simple, as the order of the inserted self-energy directly determines the order of the resulting contribution to the proper-polarization function. Moreover, since the LO contribution in $r_s$ from this $m$-th order family of diagrams always corresponds to the baseline LO $r_s$ term of the entire set $\mathcal{S}_i$ in the high-density limit, subsequent sets $\mathcal{S}_i$ (for $i>1$) contributing to $\Pi^{xc}_{EH}(q,0)$ are anchored by higher-order members of this family; this guarantees a strictly higher-order $r_s$ contribution than the previous set. As we will show, extracting the LO $r_s$ term of the $q^2$ coefficient from the $m$-th order diagrams in Fig.~\ref{family_polarization_selfenergy} bypasses the need to analyze the mathematical structure of the integrand of any other diagram in the set $\mathcal{S}_i$.

We first give the integral expressions of both diagrams illustrated in Fig.~\ref{family_polarization_selfenergy} and the sum of the two. These are given by the following
\begin{align}
\Pi^{(m)}(q,0) &= \Pi^{(m)}_{+}(q,0) + \Pi^{(m)}_{-}(q,0),\label{Pi_m_sum}\\
\Pi^{(m)}_{\pm}(q,0) &= -2i\int \frac{d^4p}{(2\pi)^4}\Sigma^{(m)}(p^{\mu})\left[G^0(p^{\mu}) \right]^2G^0(l^{\mu}_{\pm}),\label{Pi_pm}  
\end{align}
where $l^{\mu}_{\pm}$ is a shorthand notation for $l^{\mu}_{\pm}=(\vec{p}\pm\vec{q},p^0)$.
The extraction of the $q^2$ coefficient from the sum of the two diagrams follows precisely the same mathematical steps  used in our previous work for the $\Pi^{xc}_2(q,0)$ function\cite{benites2026}. In that work, we demonstrated that
the square of the Green's function can be conveniently re-expressed  as follows
\begin{align}
\left[G^0(p^{\mu})\right]^2 &= -G^0_2(p^{\mu})+2\pi i\delta(p^0)\delta(\mu_0-\epsilon^0_p), 
\label{Greens_function_squared}\\
G^0_{n}(p^{\mu}) &= \frac{\partial^{n-1} G^0(p^{\mu})}{\partial \mu_0^{n-1}},\label{G_0_n} 
\end{align}
where $G^0_{n}(p^{\mu})$ is the same notation we used in the Appendix section of Ref.~\cite{benites2026}. 

We then perform a Taylor expansion of the term $G^0(l^{\mu})$ in the $\vec{q} \to 0$ limit, where we have previously demonstrated that it can be expressed in terms of partial derivatives of the non-interacting Green's function with respect to the non-interacting chemical potential $\mu_0$. We have the following expression
\begin{align}
G^0(l^{\mu}_{\pm}) &= G^0(p^{\mu}_{\vec{\delta}}) -\left(\pm\frac{\vec{q}\cdot \vec{p}}{m} +\frac{q^2}{2m}  \right)G^0_2(p^{\mu}_{\vec{\delta}})\nonumber\\
&+ \frac{(\vec{q}\cdot \vec{p})^2}{2m^2}G^0_3(p^{\mu}_{\vec{\delta}}),
\label{Greens_function_Taylor}
\end{align}
where the $\vec{\delta} \to 0$ limit is taken when extracting the $q^2$ coefficient of $\Pi^{(m)}(q,0)$ and $p^{\mu}_{\vec{\delta}}=p^{\mu}+\vec{\delta}$. By using Eq.~\ref{Greens_function_squared} and Eq.~\ref{Greens_function_Taylor} in Eq.~\ref{Pi_pm}, and exploiting the azimuthal symmetry at the integration level, we obtain the following expression for the $q^2$ coefficient ($b^{(m)}_{xc}$) of $\Pi^{(m)}(q,0)$
\begin{align}
b^{(m)}_{xc} &= b^{(m)}_{xc,1} - \frac{m^2 \Sigma^{(m)}(k_F,0)}{12 \pi^2 k_F^3},\label{b_m_final_expression}\\
b^{(m)}_{xc,1} &= \frac{i}{m}\int\frac{d^4p}{(2\pi)^4} \Sigma^{(m)}(p^{\mu})\left[\frac{1}{3}G^0_4(p^{\mu})-\frac{\epsilon^0_p}{9}G^0_5(p^{\mu}) \right].\label{b_m_xc_1_final_expression}
\end{align}
The compact expression for $b^{(m)}_{xc,1}$ was found by using the following identity regarding the partial derivatives of the non-interacting Green's function with respect to the non-interacting chemical potential $\mu_0$. This identity is given
\begin{equation}
\lim_{\vec{\delta} \to 0} G^0_{n_1}(p^{\mu}_{\vec{\delta}})G^0_{n_2}(p^{\mu}) = -B(n_1,n_2)G^0_{n_1+n_2+1}(p^{\mu}), 
\label{identity_partial_Greens_function}
\end{equation}
where $B(n_1,n_2)$ is the well-known Beta function, with $n_1,n_2 \in \mathbb{N}$.

Because the last term in Eq.~\ref{b_m_final_expression} hosts the LO \(r_{s}\) contribution to \(b^{(m)}_{xc}\) within the set of diagrams $\mathcal{S}_i$, one only needs to determine the expanded form of the \(m\)-th order self-energy \(\Sigma^{(m)}(k_F,0)\) evaluated at the Fermi surface, \(p=k_F\). Within this topological class of self-energy insertions, the diagrammatic set index maps directly to the order of the self-energy diagram within the RPARPE, $i=m$. At first order ($m=1$), the self-energy inserted into the fermionic lines is the GW self-energy, which can be separated into two terms as illustrated in Fig.~\ref{self_GW_fig}. The first is the Fock self-energy, which scales linearly with \(k_{F}\) as established in our prior work\cite{benites2026}, dictating that this exchange contribution to \(b_{x}\) takes the characteristic \(r_{s}^{2}\) form. The other term in the separation of the GW self-energy corresponds to a resummation of the bare polarization \(\Pi_0(q,0)\) as a geometric series within the RPA, yielding the ring-like self-energy series \(\Sigma_r(k_F,0)\). In this work, we show that its leading-order \(r_{s}\) behavior scales as \(\ln r_s\). Consequently, due to the \(1/k_F^3\) prefactor in the final term of Eq.~\ref{b_m_final_expression}, its corresponding correlation contribution to \(b_{c}\) scales as \(r_s^3 \ln r_s\). Importantly, the combination of these two self-energies yields the standard GW self-energy, which serves as the perturbative insertion into the fermionic propagator lines shown in Fig.~\ref{family_polarization_selfenergy}. The proper-polarization functions generated by these insertions belong to the same diagrammatic class that produces the LO \(r_{s}^{2}\) contribution to \(b_{xc}\). The next-order diagram within the RPARPE ($m=2$), which belongs to the set $\mathcal{S}_2$, corresponds to the kite-like self-energy series \(\Sigma_{2b}(k_F,0)\) at \(p=k_F\). As derived in Appendix~\ref{Kite_like_self_energy_series}, the leading contribution of this series in \(r_{s}\) is a constant, which manifests in \(b_{xc}\) as an \(r_{s}^{3}\) correction. Because any higher-order self-energy diagram beyond \(\Sigma_{2b}(k_F,0)\) generates strictly higher-order \(r_{s}\) dependencies in the high-density limit and belongs to a set $\mathcal{S}_i$ with $i>2$, we can safely rule out any further diagrammatic corrections to the \(r_s^3 \ln r_s\) scaling calculated herein. Additional technical details regarding the extraction of the leading-order \(r_{s}\) terms for the kite-like self-energy series are cataloged in Appendix~\ref{Kite_like_self_energy_series}.

\section{Summary and Conclusions}
\label{Discussion}
In our previous work \cite{benites2026}, we calculated the leading $r_s$  contribution to the coefficient $B_{xc}$ of the $|\nabla n(\vec r)|^2$ term in the
gradient expansion approximation (GEA), resolving previous misconceptions and mathematical errors in the literature. 
Historically, all past studies--including Ref.~\onlinecite{benites2026}-- focused exclusively on evaluating the leading-order $r_s$ contribution to $b_{xc}$ (and $B_{xc}$). In the present paper, we extend our technique to calculate the next-to-leading order correction.

First, the entire exchange contribution\cite{benites2026} to $b_{xc}$ (the $b_x$ term) is given as
\begin{eqnarray}
  b_{x} &=& b^{\prime (\mathrm{LO})}_c R_+(\beta(k_F)),
  \label{bx}
\end{eqnarray}
where the factor $R_{\pm}(\beta(k_F))$ is defined as
\begin{eqnarray}
  R_{\pm}(\beta(k_F)) &\equiv& 1\pm2k_F\frac{\partial}{\partial k_F}\ln(\beta(k_F)).
  \label{Rpm}
\end{eqnarray}

Second, for the calculation of  $b_c$,
we divided
the interval of integration into two regions:
Region 1 spans the interval $x \in (0,1/2)$ and Region 2 spans  $x \in (1/2,\infty)$. We expressed the various contributions to $b_c$ given in Eq.~\ref{b_c_def} in terms of the integrals $L^j_{im}(r_s)$ (for $j=1,2$, $i=1,\dots,6$ and $m \geq 1$, which are defined by Eqs.~\ref{L1_im_structures} and~\ref{L2})
as shown in Eqs.~\ref{b'_c11_Lform}--\ref{b'''_xc_sum} and Eqs.~\ref{b'''_xc_11_Lform}--\ref{b'''_xc_16_Lform}; the decomposition of these integral kinds yields every term of the $r_s$ expansion in the $r_s \to 0$ limit. We found that the integrals in Region $2$
yield exclusively $\mathcal{D}$-type contributions in the small-$r_s$ expansion of the coefficients contributing to $b_c$. In Region $1$,
however, these integrals can be written in terms of the functions $\Gamma^{n}_{lm}(r_s,y)$, defined by the integral expression in Eq.~\ref{Gamma_n_l_m_def}. In particular, we find that the NLO term contributing to $b_{xc}$ has the form of $r_s^3 \ln(r_s)$. Specifically, the contributions from $b^{\prime}_c$, $b^r_c$ and $b^{\prime \prime \prime}_{c}$ furnish the $\mathcal{D}$-type NLO $r_s$ terms, whereas $b^{\prime \prime}_c$ yields $\mathcal{D}$-type and $\mathcal{H}$-type NLO contributions.

The final result for the  expansions of all the NLO contributions to $b_{c}$
can be summarized as
\begin{align}
b^{\prime}_{c} &\approx b^{\prime (\mathrm{LO})}_c R_-(\beta(k_F)) + C^{\prime}_{L3} r^3_s \ln(r_s) + O(r_s^3),
\label{b'_c_final_NLO_result}\\
b^{\prime \prime}_{c} &\approx b^{\prime \prime (\mathrm{LO})}_c + C^{\prime \prime}_{L3} r_s^3\ln(r_s) + O(r_s^3) ,\label{b''_c_NLO_final_result}\\
b^{\prime \prime \prime}_{c} &\approx b^{\prime \prime \prime (\mathrm{LO})}_{c} + C^{\prime \prime \prime}_{L3} r_s^3 \ln(r_s) + O(r_s^3) ,\label{b'''_xc_NLO_final_result}\\
b^r_c &\approx C^{r}_{L3} r^3_s \ln(r_s) + O(r_s^3), \label{b_r_NLO_final_result}
\end{align}
where the factor
$R_-(\beta(k_F))$ is given by Eq.~\ref{Rpm} and gives the regulator dependence. Crucially, the regulator-dependent part of $b_x$ cancels the regulator dependence of the LO contribution to $b'_c$; namely, when adding
$b_{x}+b'_c$, the combined LO result equals twice the value of
what we define as $b^{\prime (\mathrm{LO})}_c$, which is completely regulator-independent.
In past controversies resolved in Ref.~\onlinecite{benites2026},
the exchange contribution $b_x$ and correlation contribution $b_c$ were
evaluated separately by various authors who implicitly or explicitly
introduced their own choice of regulator. However, as demonstrated in Ref.~\cite{benites2026} and summarized above,
these separate contributions do not exist independently; rather, their
isolated values depend inherently on the specific regulator employed.
This dependence is explicitly reflected by the presence of $R_-(\beta(k_F))$ and $R_+(\beta(k_F))$ in the isolated expressions for $b_x$ and $b_c$, respectively.

When combined, however, their dependencies cancel via
the identity $R_-(\beta(k_F))+R_+(\beta(k_F))=2$, and the regulator
dependence completely disappears.
Consequently, evaluating the total sum $b_{xc} = b_x + b_c$ under a single,
consistent regulator yields a combined result that is entirely invariant.
This cancellation is physically expected because the physical response of an interacting electron gas to an external inhomogeneous electric field is determined by $b_{xc}$ and not by $b_x$ or $b_c$ individually; this standard breakup serves purely as a conceptual and
computational tool. This subtle point was overlooked in previous
literature (with the exception of Ref.~\onlinecite{benites2026}),
which propagated the fundamentally incorrect narrative that $b_x$ and $b_c$
possess independently well-defined, regulator-invariant values.
As a result, older calculations introduced severe errors by combining
values of $b_x$ and $b_c$ obtained under mismatched regulators. If we take, for example,
two different regulator $\beta(k_F)$ and $\beta^{\prime}(k_F)$ to employ in
the calculation of 
$b_x$ and $b_c$ respectively, the resulting sum 
$R_-(\beta(k_F))+R_+(\beta'(k_F))$ fails to cancel the regulator dependence.
This, instead
yields un-physical, nonsensical values.

Furthermore, as seen in Eq.~\ref{bx}, there is no further $r_s$ contribution to $b_x$.
The values of $b^{\prime (\mathrm{LO})}_c$, $b^{\prime \prime (\mathrm{LO})}_c$, $b^{\prime \prime \prime (\mathrm{LO})}_{c}$, and $b^r_c$ are tabulated in Table~\ref{table_b_c_MB}, while the coefficients of the $r_s^3 \ln(r_s)$ term are listed in Table~\ref{table_coeff_NLO}.

\setlength{\tabcolsep}{4pt}
\def\arraystretch{1.5}
\begin{table}[ht]
    \centering
    \begin{tabular}{|c|c|c|}
        \hline
         $b^{\prime (\mathrm{LO})}_c$& $b^{\prime \prime (\mathrm{LO})}_c$ &$b^{\prime \prime \prime (\mathrm{LO})}_{c}$ \\
        \hline
         $5/72$& $0.10359$ &$0.07392$  \\
        \hline
    \end{tabular}
    \caption{Values of the coefficients that contributes as $r_s^2$ to $b_c$ obtained in leading order in $r_s$. These reported values are given in units of a common factor of $\tau = e^2m^2/\pi^3k_F^2$.}
    \label{table_b_c_MB}
\end{table}

\setlength{\tabcolsep}{4pt}
\def\arraystretch{1.5}
\begin{table*}[ht]
    \centering
    \begin{tabular}{|c|c|c|c|}
        \hline
          $C^{\prime}_{L3}$ & $C^{\prime \prime}_{L3}$  & $C^{\prime \prime \prime}_{L3}$ & $C^{r}_{L3}$ \\
        \hline
         $301/3456-\ln(2)/6$ & $(32\ln(2)-13)/384$ & $-1/288$ & $(\ln(2)-1)/12$ \\
        \hline
    \end{tabular}
    \caption{The coefficients of the $r^3_s \ln(r_s)$ terms corresponding to $b_c$ found by using our systematic approach. These values in the table are given in units of a common factor of $\tau_2 = (m \alpha a_B)^3e^4/\pi^4$.}
    \label{table_coeff_NLO}
\end{table*} 

By using Eqs.~\ref{b'_c_final_NLO_result}--\ref{b_r_NLO_final_result}, we find that
\begin{eqnarray}
  b_{xc} & = &  b^{(\mathrm{LO})}_{xc}  + C_{L3} r^3_s \ln(r_s) +
  O(r_s^3) + ...,\\
  C_{L3} &=& -\frac{29}{1944}\frac{e^4 (m a_B)^3}{\pi^5},\\
  b^{(\mathrm{LO})}_{xc} &=& 2b^{\prime (\mathrm{LO})}_c+b^{\prime \prime (\mathrm{LO})}_c+b^{\prime \prime \prime (\mathrm{LO})}_c,
\end{eqnarray}
and combining this with the expression for $B_{xc}[n]$
given by Eq.~\ref{B_coeff}, we obtain
\begin{equation}
B_{xc}[n] = B^{MB-S}_{xc} - \frac{29}{3888 \alpha^3 \pi^4} \frac{\ln(r_s)}{e^2 m^2 a_B^6 r_s^3},
\label{B_xc_NLO}    
\end{equation}
which is the same as the $B_{xc}[n]$ given in Eq.~\ref{Bxc_final} when using atomic units. This provides a constraint to be implemented in a GGA functional in the $r_s \to 0$ limit. 

One of our primary conclusions is that the diagrams illustrated in Figs.~\ref{GA_Fock} and~\ref{Pi_c}, which contribute to $\Pi^{xc}(q,0)$, are the exclusive terms that contribute to the coefficient of $r_s^3\ln(r_s)$
of $b_{xc}$ in the long-wavelength and high-density limits.
Consequently, the combination of the coefficients
$C^{\prime}_{L3}$, $C^{\prime\prime}_{L3}$, $C^{\prime\prime\prime}_{L3}$, and $C^r_{L3}$ is
exact and meaningful, as their combined result $C_{L3}$ receives no corrections from other higher-order diagrams within RPARPE. They uniquely determine the coefficient of $r_s^{-3} \ln r_s$ of $B_{xc}[n]$, providing an exact constraint for generalized gradient approximation (GGA) functionals in the limits of
small $r_s$ and smooth density fluctuations.

However, obtaining the full contribution to the coefficient of the next-to-next-to-leading-order (NNLO) term---namely, the coefficient of the $r_s^3$ term
in $b_{xc}$ (See Eqs.~\ref{b'_c_final_NLO_result}--\ref{b_r_NLO_final_result})---requires calculating higher-order proper-polarization diagrams beyond those illustrated in Figs.~\ref{GA_Fock} and~\ref{Pi_c}.
This task would necessitate going beyond the leading order of RPARPE.
If one intends to evaluate the coefficient of the  $r^3_s$ terms
contributing to $b_c$ while remaining strictly within our
leading-order RPARPE framework
(Figs.~\ref{GA_Fock} and~\ref{Pi_c}), it would be necessary
to incorporate both the $\mathcal{D}$-type and $\mathcal{H}$-type
contributions discussed in Appendix~\ref{different_contribution_types_appendix}. However, the resulting
coefficient obtained via this approach would be incomplete.
This incompleteness arises because diagrams
from non-leading order in our RPARPE---which are
omitted from Figs.~\ref{GA_Fock} and~\ref{Pi_c}---also contribute to the coefficient of the $r^3_s$
order term in $b_{xc}$.

If one intends to calculate the next-order gradient term, as defined by Eq.~\ref{dimensionless_gradient}, within the GEA, it would be necessary to evaluate
the fourth-order exchange-correlation kernel $K^{(4)}_{xc}(q,0)$.
This kernel can be obtained by taking a second functional derivative of the second order exchange-correlation kernel $K_{xc}(q,0)$ given in Eq.~\ref{B_xc_wavevector}. It is straightforward to demonstrate that the $q^4$ term from the expansion in the $q \to 0$ limit of $K^{(4)}_{xc}(q,0)$ yields the
$|\nabla n(\vec{r})|^4$ term of the gradient expansion approximation. The 
coefficient of this term scales as $r_s^{12}$, which implies
that the LO in $r_s$ of the coefficient of the $s^4$ term
scales as $r_s^{-4}$, matching the LO behavior of $B_{xc}$.   

However, the next-order term to be appended to the exchange-correlation functional in the slowly varying density limit is not of the form $|\nabla n(\vec{r})|^4$; rather, it corresponds to the square of the Laplacian of the density, $(\nabla^2 n(\vec{r}))^2$. It is straightforward to demonstrate that this term of the GEA is captured by the $q^4$ term of the expansion of the $K_{xc}(q,0)$ kernel in the $q \to 0$ limit. By dimensional analysis, we can predict that the LO in $r_s$ in the high-density limit of this $q^4$ term from $K_{xc}(q,0)$
scales as $r_s^6$. This $r_s$ scaling behavior agrees with the $q^4$ coefficient expression from the expanded second-order exchange kernel in Ref.~\onlinecite{PhysRevB.54.17402}.    

In conclusion, our calculated value of the NLO coefficient in $r_s$---corresponding to the $r_s \ln r_s$ term relative to the leading order---receives no corrections from other higher-order diagrams omitted from the set $\mathcal{S}_1$. It is, therefore,
an exact and definite value to which any GGA functional should be
constrained in the limits of $r_s \to 0$ and slowly varying density.
Conversely, the exact coefficient of the $s^4$ term for use in GGA functionals
remains undetermined.
Evaluating all contributions to this higher-order coefficient requires going beyond the leading order in the RPARPE framework, as it cannot be obtained simply by expanding
the diagrammatic expressions shown in Figs.~\ref{GA_Fock} and~\ref{Pi_c}.

\section{acknowledgments}
  This work was supported by the U.S. National Science Foundation under Grant No. NSF-DMR-2110814.
\appendix
\section{$\Gamma^{2n+1}_{lm}(r_s,y)$ integral kind}
\label{Gamma_integral_kinds_section}
This appendix section explains how the integral kind $\Gamma^{2n+1}_{lm}(r_s,y)$ are expanded in terms of the Wigner-Seitz radius $r_s$ (for any $n \in \mathbb{Z}^+$) in the high-density limit ($r_s \to 0$). We start from the definition of these functions given by Eq.~\ref{Gamma_n_l_m_def}. Performing the change of variable $u = \Delta_1(x,y)$, we obtain the following expression
\begin{widetext}
\begin{align}
\Gamma^{2n+1}_{lm}(r_s,y) = \frac{(l+m-1)!}{2l!(m-1)!\gamma^{n+1}(r_s,y)} \sum^{n}_{l_3=0} &\int^{u_2}_{u_1}  du \frac{n! u^{l_3-l-m}[\Delta_3(u,y)]^{l}}{l_3!(n-l_3)!}\left[-\frac{\alpha r_s}{4 \pi}Q(0,y)\right]^{n-l_3},
\label{Gamma_function_n_odd_u}\\
    \Delta_3(u.y) = \sum^{\infty}_{n=2} \frac{Q^{(2n,0)}(0,y)}{\gamma^{n}(r_s,y)(2n)!}\left(u-\frac{\alpha r_s Q(0,y)}{4 \pi}\right)^n,\label{Delta3}
\end{align}
\end{widetext}
for $n \in \mathds{Z}^+$ ,where $u_1$ ($u_2$) is a shorthand notation for $u_1 = \Delta_1(0,y)$ [$u_2= \Delta_1(\frac{1}{2},y)$]. The integral over the $u$ variable is straightforward to calculate, but we must isolate the special case $l=0$ and $l_3-m=-1$, which yields a logarithmic term upon integration. For these specific values of $l$ and $l_3$, this generates terms with the form $r^{n-m+1}_s \ln(r_s)/\gamma^{n+1}(r_s,y)$. This specific $r_s$ dependence originates from the integral expressions for the $b^{'}_c$, $b^{''}_c$, and $b^{'''}_{c}$ coefficients, as well as the self-energy term $\Sigma_r(k_F,0)$, since they contain $\Gamma^{2n+1}_{0m}(r_s,y)$ factors. In addition, the prefactor $\gamma^{-(n+1)}(r_s,y)$ can be expanded in the $r_s \to 0$ limit as a geometric series. This expansion generates higher-order $r_s$ terms beyond $r^{n-m+1}_s \ln(r_s)$, which must be systematically accounted for when determining the complete coefficient associated with this logarithmic power of $r_s$. 

For the remaining terms in the series in Eq.~\ref{Gamma_function_n_odd_u} that do not satisfy $l_3-m=-1$, integrating over $u$ yields terms of the power-law form $r_s^{n-m+1}$. We find that the leading contribution in $r_s$ to $b^{'}_c$,$b^{''}_c$, $b^{'''}_{c}$, and $\Sigma_r(k_F,0)$ stems from the condition $l_3-m\neq-1$. In general, we can conclude that $\Gamma^{2n+1}_{lm}(r_s,y)$ decomposes as
\begin{widetext}
\begin{eqnarray}
    \Gamma^{2n+1}_{lm}(r_s,y) &=& \sum^{\infty}_{j=n+l+1-m} [\delta_{l+n\geq m-1}F^{n,l,m}_{Lj}(y)r^j_s\ln(r_s)+F^{n,l,m}_j(y)r_s^j],\label{Gamma_decomposition_rs}
\end{eqnarray}
\end{widetext}
where it is straightforward to find a closed expression for the function $F^{n,l,m}_{L,n+l+1-m}(y)$, which captures the leading contribution in $r_s$ of $\Gamma^{2n+1}_{lm}(r_s,y)$ and is of the logarithm type; this contributes to the NLO $r_s$ term to the $b_c$ coefficient. The closed expression of this function is given by
\begin{widetext}
\begin{equation}
F^{n,l,m}_{L,n+l+1-m}(y) = \frac{(-1)^{l+n-m} (2l+n)!\left[Q^{(4,0)}(0,y) \right]^l }{2l!(m-1)!(l+n-m+1)!}\left[\frac{\alpha}{4 \pi}Q(0,y)\right]^{n+l+1-m},
\label{F^nlm_L,n+1-m}    
\end{equation}
\end{widetext}
while the remaining functions $F^{n,l,m}_{j}(y)$ and $F^{n,l,m}_{L,j>n+l+1-m}(y)$ are not needed to determine the NLO term in $r_s$ in this work, though they could be systematically evaluated by substituting Eq.~\ref{Delta3} into Eq.~\ref{Gamma_function_n_odd_u}. 

\section{Different contribution types to the coefficients of the $r_s$ terms}
\label{different_contribution_types_appendix}
We classify the different contributions to the coefficients of the $r_s$ terms (originating from the $r_s$-expansion of every term contributing to $b_c$ through Eq.~\ref{b_c_def} in the $r_s \to 0$ limit) into two distinct types: $\mathcal{D}$ (stands for Direct) and $\mathcal{H}$ (stands for Hybrid). As we discuss in this section, these two types arise only in the integral kinds $L^1_{im}(r_s)$, whereas $L^2_{im}(r_s)$ contributes exclusively as $\mathcal{D}$-type. Below, we first discuss the $\mathcal{D}$-type contributions to the $r_s$ coefficients, followed by the $\mathcal{H}$-type contributions.
\subsection{$D$-type contribution}
\label{D-type}
In this subsection, we discuss the $\mathcal{D}$-type contribution to every coefficient of the terms arising from the $r_s \to 0$ expansion that originates from both $b_c$-primed coefficients, $b^{'''}_{c}$, and $\Sigma_r(k_F,0)$. $\mathcal{D}$-type contributions can be further categorized into $\mathcal{D}_1$ and $\mathcal{D}_2$ sub-types. After applying the first two steps of our systematic approach, $\mathcal{D}_1$-type coefficients are defined as the contributions obtained by making the approximations $\bar{\epsilon}(x,y)\approx \Delta_1(x,y)$ and $\gamma(r_s,y)\approx 1$ in the denominator of the integral expressions for the $\lambda$-independent contributions to $b_c$ in Region $1$ of integration [$x \in (0,1/2]$]. This implies that only the leading term in the Taylor expansion of $Q(x,y)$ is retained in the denominator at the integrand level. $\mathcal{D}_2$-type contributions, however, capture every single term in the $r_s \to 0$ expansion originating from the second region of integration [$x \in (1/2,\infty)$]. This means that the full expansion in $r_s$ of the integral kind $L^2_{lm}(r_s)$, given by Eq.~\ref{L2_im_expansion_rs}, is classified as a $\mathcal{D}_2$-type contribution. These are considered Direct-type contributions because obtaining the $r_s$ expansion in Region $2$ is straightforward, as a subdivision into sub-regions of integration is not required to determine the $r_s$ dependence of this integral kind, and their two-dimensional integrals can be easily computed numerically.

We can obtain the $\mathcal{D}_1$-type terms of the $r_s$ expansion by using the integral kind $\Gamma^{2n+1}_{0m}(r_s,y)$, given by the following expression
\begin{widetext}
\begin{equation}
\Gamma^{2n+1 (\mathcal{D}_1)}_{0m}(r_s,y) = \frac{1}{2}\sum^{n}_{l_3=0}\left(-\frac{\alpha r_s}{4 \pi}Q(0,y) \right)^{n-l_3} \int^{u_2}_{u_1}du \frac{n!u^{l_3-m}}{l_3!(n-l_3)!},
\label{Gamma^n_0m}   
\end{equation}
\end{widetext}
where the result of the integral over the $u$ variable is separated into two terms as follows
\begin{widetext}
\begin{align}
\Gamma^{2n+1 (\mathcal{D}_1)}_{0m}(r_s,y) =& \delta_{n \geq m-1}\Gamma^{2n+1}_{0m,1}(r_s,y)+\Gamma^{2n+1}_{0m,2}(r_s,y),
\label{Gamma^n_0m_integral_u}\\
\Gamma^{2n+1 (\mathcal{D}_1)}_{0m,1}(r_s,y) =& \frac{1}{2}\frac{n!}{(m-1)!(n+1-m)!}\left(-\frac{\alpha r_s}{4 \pi}Q(0,y) \right)^{n+1-m} \ln\left|\frac{1+\frac{\alpha r_s}{\pi}Q(0,y)}{\frac{\alpha r_s}{\pi}Q(0,y)} \right|,\label{Gamma^n_0m,1}\\
\Gamma^{2n+1 (\mathcal{D}_1)}_{0m,2}(r_s,y) =& \frac{1}{2} \sum^{n}_{l_3 \neq m-1}\frac{n!}{(l_3)!(n-l_3)!(l_3+1-m)}\left(-\frac{\alpha r_s}{4 \pi}Q(0,y) \right)^{n-l_3} \Delta^{l_3,m}_3(r_s,y),\label{Gamma^n_0m,2}\\
 \Delta^{l_3,m}_3(r_s,y) =& \left[\frac{1}{4}+\frac{\alpha r_s Q(0,y)}{4\pi}\right]^{l_3+1-m}-\left[\frac{\alpha r_s Q(0,y)}{4 \pi} \right]^{l_3+1-m},
\end{align}
\end{widetext}
where the sum over the index $l_3$ runs from $l_3=0$ up to $l_3=n$. Utilizing these expressions yields the $\mathcal{D}_1$-type contributions as follows
\begin{align}
    \Gamma^{2n+1 (\mathcal{D}_1)}_{0m}(r_s,y) =&  \delta_{n \geq m-1} F^{n,m (\mathcal{D}_1)}_{L,n+1-m}(y)r^{n+1-m}_s\ln(r_s) \nonumber \\
    & + \sum^{\infty}_{j=n+1-m}F^{n,m (\mathcal{D}_1)}_j(y)r_s^j,\label{Gamma_decomposition_rs_D1}
\end{align}
where the function $F^{n,m (\mathcal{D}_1)}_{L,n+1-m}(y)$ is given by the following expression
\begin{equation}
F^{n,m  (\mathcal{D}_1)}_{L,n+1-m}(y) = F^{n,0,m}_{L,n+1-m}(y),
\label{F^nlmD1_L,n+1-m}    
\end{equation}
which implies that the leading term in $r_s$ of the form $r^{n+1-m}_s \ln(r_s)$ originating from $\Gamma^{2n+1}_{0m}(r_s,y)$ is a $\mathcal{D}_1$-type contribution. Although the remaining coefficients $F^{n,m (\mathcal{D}_1)}_j$ do not possess a compact closed form like $F^{n,m (\mathcal{D}_1)}_{L,n+1-m}$, they can be determined systematically by performing a Taylor expansion in the $r_s \to 0$ limit of the integral kinds $\Gamma^{2n+1}_{0m,1}(r_s,y)$ and $\Gamma^{2n+1}_{0m,2}(r_s,y)$.

The $\mathcal{D}_1$-type contributions from the $r_s$-expansion of the integral kind $\Gamma^{2n+1}_{0m}(r_s,y)$ given by Eq.~\ref{Gamma_decomposition_rs_D1} can be substituted into the expressions for the integral kind $L^1_{im}(r_s,y)$ as given by Eq.~\ref{L1_im_l=0}. This yields the following $\mathcal{D}_1$ contributions to the expansion in $r_s$ of the integral kind $L^1_{im}(r_s,y)$
\begin{align}
    L^{1 (\mathcal{D}_1)}_{im}(r_s) =& \sum^{\infty}_{n=0} \delta_{n \geq m-1} C^{i (\mathcal{D}_1)}_{L, n+1-m} r_s^{n+1-m} \ln(r_s) \nonumber \\
    &+ \sum^{\infty}_{n=0} \sum^{\infty}_{j=n+1-m}C^{i (\mathcal{D}_1)}_{n,m,j} r_s^j,
\label{L1_im_final_rs_decomposition}
\end{align}
where $m \geq 1$, the $\mathcal{D}_1$-type contribution to the coefficients of the $r^{n+1-m}_s \ln(r_s)$ and $r_s^j$ terms are given by
\begin{align}
C^{i (\mathcal{D}_1)}_{L,n+1-m} =& \int^{\infty}_0 dy \frac{f^{(2n+1,0)}_i(0,y)}{(2n+1)!}F^{n,m (\mathcal{D}_1)}_{L,n+1-m}(y),
\label{C^iD1_L,n+1-m_final_integrals}\\
C^{i (\mathcal{D}_1)}_{n,m,j} =& \int^{\infty}_0 dy \frac{f^{(2n+1,0)}_i(0,y)}{(2n+1)!}F^{n,m (\mathcal{D}_1)}_{j}(y).\label{C^iD1_nmj_final_integral}
\end{align}
The leading term in $r_s$ for the $\mathcal{D}_1$-type contribution to the integral kind $L^{1 (\mathcal{D}_1)}_{im}(r_s)$ is given by $r^{n+1-m}_s \ln(r_s)$, which occurs only for $n \geq m-1$. 

\subsection{$\mathcal{H}$-type contribution}
We discuss the $\mathcal{H}$-type contributions to every coefficient in the $r_s$-expansion, which arise exclusively from the integral kinds $L^1_{im}(r_s)$, since Region $2$ provides only $\mathcal{D}$-type contributions. These hybrid-type contributions arise by taking into account higher-order terms in the Taylor expansion of $Q(x,y)$ in the $x \to 0$ limit. Consequently, we have $\gamma(r_s,y) \not \approx 1$, a condition that yields higher-order $r_s$ contributions beyond the dominant logarithmic term extracted from the $\mathcal{D}_1$-type contributions to $L^1_{im}(r_s)$ from the integral kinds $\Gamma^{2n+1}_{0m}(r_s,y)$. Another origin of higher-order terms in $r_s$ is the function $\Delta_3(u,y)$ given by Eq.~\ref{Delta3}.

Let us discuss the two main $\mathcal{H}$-type contributions to the $r_s^j \ln(r_s)$ and $r_s^j$ terms. One primary $\mathcal{H}$-type contribution in $r_s$ arises from the Taylor expansion of the denominator term
\begin{flalign}
\frac{1}{[\gamma(r_s,y)]^n} \!=\! \sum^{\infty}_{l=0} \frac{(l+n-1)!}{l!(n-1)!}\left[-\frac{\alpha r_s}{8 \pi}Q^{(2,0)}(0,y) \right]^{l}\!, \!\!\!\!\!&&
\label{reciprocal_gamma_n}    
\end{flalign}
which originates from the expressions for $\Delta_3(r_s,y)$ and the integral kind $\Gamma^{2n+1}_{lm}(r_s,y)$ given by Eq.~\ref{Delta3} and Eq.~\ref{Gamma_function_n_odd_u}, respectively. For instance, including the $r_s$ terms coming specifically from the integral kind $\Gamma^{2n+1}_{0m}(r_s,y)$ yields a prefactor of $1/\gamma^{n+1}(r_s,y)$ in Eq.~\ref{Gamma^n_0m,1}. In combination with Eq.~\ref{reciprocal_gamma_n}, this generates higher-order contributions beyond the logarithmic term $r_s^{n+1-m}\ln(r_s)$ (under the condition $n\geq m-1$). Similarly, this prefactor gives rise to terms of higher-order than $r_s^{n+1-m}$.

The other $\mathcal{H}$-type contribution to the $r_s$ terms arising from $L^1_{lm}(r_s)$ originates from the function $[\Delta_3(u,y)]^l$ contained in Eq.~\ref{Gamma_function_n_odd_u}. The inclusion of this function yields higher-order $r_s$ terms of the form $r_s^{n+l+1-m}\ln(r_s)$ and $r_s^{n+l+1-m}$.

\section{List of functions used in this work}
\label{list_functions_used}
We list the expressions for the functions used throughout our calculations
\begin{align}
    C(0,y) =& \frac{y^2(13+9 y^2)}{36 (1+y^2)^3},
    \label{C_0}\\
    Q(0,y) =& 4\left[1-y\tan^{-1}\left(\frac{1}{y}\right) \right],\label{Q_0}\\
    I^{\prime}_2(0,y) =& -\left(\frac{3}{4}\right)\frac{y^2(y^2-3)}{(1+y^2)^3},\label{I2_prime_0}\\
    h_1(0,y) =& \frac{2y^2+1}{(1+y^2)^2},\label{h1_0}\\
    h_2(0,y) =& \frac{2}{1+y^2},\label{h2_0}\\
    h_3(0,y) =& \frac{4}{(1+y^2)^2}.\label{h3_0}
\end{align}
The partial derivatives of the functions required to obtain the NLO $r_s$ term contributing to the $b_{xc}$ coefficient are given below
\begin{align}
C^{(2,0)}(0,y) =& \frac{12 y^8+9y^6-35y^4-9y^2-9}{18(1+y^2)^5}, 
\label{C_2_0}\\
I^{\prime (2,0)}_{2}(0,y) =& \frac{2y^8+25y^6-55y^4+15y^2-3}{2(1+y^2)^5},\label{I2_prime_20}\\
Q^{(2,0)}(0,y) =& -\left(\frac{8}{3}\right) \frac{1}{(1+y^2)^2},\label{Q_2_0}\\
g^{(1,0)}(0,y) =& \frac{4}{1+y^2},\label{g_10}\\
g^{(3,0)}(0,y) =& -\frac{8(3y^2-1)}{(1+y^2)^3},\label{g_30}\\
h^{(2,0)}_1(0,y) =& -\left(\frac{4}{3} \right) \frac{(14y^4+11y^2+3)}{(1+y^2)^4},\label{h1_20}\\
h^{(4,0)}_1(0,y) =& \left(\frac{24}{5}\right) \frac{(110y^6+23y^4-12y^2-5)}{(1+y^2)^6},\label{h1_40}\\
h^{(2,0)}_2(0,y) =& -\left(\frac{4}{3}\right)\frac{(7y^2+3)}{(1+y^2)^3},\label{h2_20}\\
h^{(4,0)}_2(0,y) =& \left(\frac{16}{5}\right) \frac{55y^4+2y^2-5}{(1+y^2)^5},\label{h2_40}\\
h^{(2,0)}_3(0,y) =& -\left(\frac{16}{3}\right) \frac{(7y^2+3)}{(1+y^2)^4},\label{h3_20}\\
h^{(4,0)}_3(0,y) =& \left(\frac{32}{15}\right) \frac{575y^4+222y^2+15}{(1+y^2)^6}.\label{h3_40}
\end{align}
In this work, we kept track of the ($2n+1$)-th order of the Taylor expansion of the family of odd functions in the $x$ variable, $f_i(x,y)$ in the $x \to 0$ limit (for $n\in\mathbb{Z}^+$ and $i=1,\dots,9$), where $f_i(x, y)$ is given by
\begin{equation}
 f_i(x,y) \stackrel{x\to 0}{=} \sum^{\infty}_{n=0}\frac{x^{2n+1}}{(2n+1)!}f^{(2n+1,0)}_i(0,y).\label{f1_taylor}
\end{equation}
For the calculation of the NLO $r_s$ term contributing to $b^{'}_c$, we have used in Eq.~\ref{f1} the following expressions
\begin{align}
f^{(1,0)}_1(0,y) =& C(0,y)Q(0,y),\label{f_1_10}\\
f^{(3,0)}_1(0,y) =& 3\left[C^{(2,0)}(0,y)Q(0,y)\right. \nonumber \\
&\left.+C(0,y)Q^{(2,0)}(0,y) \right].\label{f_1_30}  
\end{align}
For the calculation of the NLO $r_s$ term contributing to $b^{\prime \prime}_c$, we have used in Eq.~\ref{f2} the following expressions
\begin{align}
 f^{(1,0)}_2(0,y) =& I^{\prime}_2(0,y),\label{f_2_10}\\
 f^{(3,0)}_2(0,y) =& 3 I^{\prime (2,0)}_2(0,y).\label{f_2_30}    
\end{align}
For the calculation of the NLO $r_s$ term contributing to $b^{r}_c$, we have used in Eq.~\ref{f3} the following expressions
\begin{align}
 f^{(1,0)}_3(0,y) =& Q(0,y)g^{(1,0)}(0,y),\label{f_3_10}\\
 f^{(3,0)}_3(0,y) =& 3Q^{(2,0)}(0,y)g^{(1,0)}(0,y) \nonumber \\
 &+Q(0,y)g^{(3,0)}(0,y).\label{f_3_30}
\end{align}
For the calculation of the coefficient of the $r_s^3 \ln(r_s)$ term contributing to $b^{\prime \prime \prime}_{c}$, we have used in Eqs.~\ref{f4}-~\ref{f9} the following expressions
\begin{align}
f^{(1,0)}_4(0,y) =& \left[g^{(1,0)}(0,y) \right]^2,\label{f_4_10}\\
f^{(3,0)}_4(0,y) =& 2\left[g^{(3,0)}(0,y)g^{(1,0)}(0,y) \right],\label{f_4_30}\\
f^{(1,0)}_5(0,y) =& 0,\label{f_5_10}\\
f^{(3,0)}_5(0,y) =& 6\left[g^{(1,0)}(0,y) \right]^2,\label{f_5_30}\\
f^{(5,0)}_5(0,y) =& 40g^{(3,0)}(0,y)g^{(1,0)}(0,y),\label{f_5_50}\\
f^{(1,0)}_6(0,y) =& f^{(3,0)}_6(0,y)=0,\label{f_6_10_and_f6_30}\\
f^{(5,0)}_6(0,y) =& 5!\left[g^{(1,0)}(0,y) \right]^2,\label{f_6_50}\\
f^{(7,0)}_6(0,y) =& \frac{7!}{3}g^{(3,0)}(0,y)g^{(1,0)}(0,y),\label{f_6_70}\\
f^{(9,0)}_6(0,y) =& 2\frac{9!}{5!}g^{(5,0)}g^{(1,0)}(0,y) \nonumber \\
&+9!\left[\frac{g^{(3,0)}(0,y)}{3!} \right]^2,\label{f_6_90}\\
f^{(1,0)}_7(x,y) =& f^{(1,0)}_4(0,y)h_1(0,y),\label{f_7_10}\\
f^{(3,0)}_7(x,y) =& 3f^{(1,0)}_4(0,y)h^{(2,0)}_1(0,y) \nonumber \\
&+f^{(3,0)}_4(0,y)h_1(0,y),\label{f_7_30}\\
f^{(1,0)}_8(0,y) =& 0,\label{f_8_10}\\
f^{(3,0)}_8(0,y) =& f^{(3,0)}_5(0,y)h_2(0,y),\label{f_8_30}\\
f^{(1,0)}_9(0,y) =& f^{(1,0)}_4(0,y)h_3(0,y).\label{f_9_10}
\end{align}

\section{Finding a two-dimensional integral expression for $\Sigma_r(k_F,0)$}
\label{Explicit_steps_Sigmar}
We first use the definition of $\Sigma_r(p^{\mu})$ to determine the $r_s$ expansion form of $\Sigma_r(k_F,0)$. By setting the frequency variable $p^0=0$ in Eq.~\ref{self-energy_ring}, we obtain the following integral expression
%
\begin{flalign}
\Sigma_r(p,0) =& i \int \frac{d^4k}{(2\pi)^4}\tilde{V}(k)\left(\frac{1}{\epsilon(k^{\mu})}-1\right)\times \nonumber \\
&\left[\frac{1}{k^0-\epsilon^0_{\vec{p}+\vec{k}}+\mu_0+i\mathrm{sign}\left(\epsilon^0_{\vec{p}+\vec{k}}-\mu_0\right)}\right], \!\!\!\!\!\!&&
\label{self_ring_step1}
\end{flalign}
%
from which we observe that the complex poles lie below (above) the real frequency axis when $\epsilon_{\vec{p}+\vec{k}}>\mu_0$ ($\epsilon_{\vec{p}+\vec{k}}<\mu_0$), as illustrated in Fig.~\ref{self_ring_poles}. We can map the integral over the real frequency axis $k^0$ (path $C_1$) into an integral over the imaginary axis (path C3) by exploiting Cauchy's theorem by doing a deformation over the contour complex path as illustrated in Fig.~\ref{self_ring_poles}, where no complex pole is enclosed by the complex contour path. After this step, we obtain the following expression
%
\begin{align}
\Sigma_r(p,0) =& -\int \frac{d^3k}{(2\pi)^3}\int^{\infty}_{-\infty} d\nu\left(\frac{1}{\epsilon(k,i\nu)}-1\right) \nonumber \\
&\times \left[\frac{\tilde{V}(k)}{i\nu-\epsilon^0_{\vec{p}+\vec{k}}+\mu_0}\right],
\label{self_ring_step2}    
\end{align}
%
which can be re-expressed in a more convenient form as follows
%
\begin{align}
\Sigma_r(p,0) =& \int \frac{d^3k}{(2\pi)^3}\int^{\infty}_{-\infty} d\nu \tilde{V}(k)\left(\frac{1}{\epsilon(k,i\nu)}-1\right) \nonumber \\
&\times\left[\frac{\epsilon^0_{\vec{p}+\vec{k}}-\mu_0+i\nu}{\nu^2+(\epsilon^0_{\vec{p}+\vec{k}}-\mu_0)^2}\right],
\label{self_ring_step3}    
\end{align}
%
where the term with $\nu$ in the numerator renders that part of the integrand an odd function in $\nu$, so the the integration over the entire imaginary frequency variable $\nu$ vanishes, simplifying the integral in Eq.~\ref{self_ring_step3} to
%
\begin{align}
\Sigma_r(p,0) =& \int \frac{d^3k}{(2\pi)^3}\int^{\infty}_{-\infty} d\nu \tilde{V}(k)\left(\frac{1}{\epsilon(k,i\nu)}-1\right) \nonumber \\
&\times\left[\frac{\epsilon^0_{\vec{p}+\vec{k}}-\mu_0}{\nu^2+(\epsilon^0_{\vec{p}+\vec{k}}-\mu_0)^2}\right].
\label{self_ring_step4}    
\end{align}
%
It is convenient to rescale the variables of integration in Eq.~\ref{self_ring_step4} as follows: we first rescale $\vec{k}\to 2k_F \vec{k}\,'$ and $\nu \to k_F^2 \nu^{'}/m$, then rescale $\nu^{'} \to k^{'}y$, and finally let $k' \to 2x$. Using this parametrization, integrating over the azimuthal and polar angles, setting $p=k_F$, and exploiting the fact that the resulting integrand is an even function in the $y$ variable, we obtain the compact two-dimensional expression
\begin{equation}
\Sigma_r(k_F,0) = -\frac{e^2}{4 \pi^3 a_B} \int^{\infty}_0 \!\!\!\!dx \int^{\infty}_0 \!\!\!\!dy \frac{Q(x,y)g(x,y)}{x^2+\frac{\alpha r_s}{4\pi}Q(x,y)}, 
\label{self_ring_kf_2d}    
\end{equation}
which can be separated into two distinct regions of integration: Region $1$, $x \in (0,1/2)$, and Region $2$, $x \in (1/2,\infty)$, with $y\in(0,\infty)$ in both cases. This decomposition yields Eqs.~\ref{Sigma_r_decomposition}--\ref{self_r2_2d}. Substituting these into Eq.~\ref{br_c}, we determine the expressions for $b^r_{c,1}$ ($b^r_{c,2}$) in terms of the integral kinds $L^1_{31}(r_s)$ ($L^2_{31}(r_s)$).

\section{Expressions of $\Gamma^n_{0m}(r_s,y)$ used in our work}
\label{Expressions_Gamma_used}
We present the analytic expressions used in this work to determine the coefficients of the leading and NLO $r_s$ terms contributing to $b_c$.
Below are the expressions for $\Gamma^{n}_{0m}(r_s,y)$ in terms of the leading $r_s$-contributions, as used for $b^{\prime}_{c,11}$, $b^{\prime \prime}_{c,11}$, $b^r_{c,1}$, and $b^{\prime \prime \prime}_{c,1}$
\begin{align}
\Gamma^{1}_{02}(r_s,y) \approx& \frac{2 \pi}{\alpha r_s Q(0,y)} + O(r_s^0),
\label{Gamma1_02_rs_expression}\\
\Gamma^{3}_{02}(r_s,y) \approx& -\frac{1}{2}\ln(r_s) + O(r_s^0),
\label{Gamma3_02_rs_expression}\\
\Gamma^{1}_{01}(r_s,y) =& \frac{1}{2\gamma(r_s,y)}\ln\left|\frac{\gamma(r_s,y)+\frac{\alpha r_s Q(0,y)}{\pi}}{\frac{\alpha r_s Q(0,y)}{\pi}}\right|,\label{Gamma1_01_rs_expression}\\
\Gamma^{3}_{01}(r_s,y) \approx& \frac{1}{8}+\frac{\alpha r_s Q(0,y)}{8 \pi} \ln(r_s) + O(r_s),\label{Gamma3_01_rs_expression}\\
\Gamma^{1}_{03}(r_s,y) \approx& \frac{1}{4}\left(\frac{4 \pi}{\alpha r_s Q(0,y)}\right)^2 + O(r_s^{-1}),\label{Gamma1_03_rs_expression}\\
\Gamma^{3}_{03}(r_s,y) \approx& \frac{\pi}{\alpha r_s Q(0,y)} + O(r_s^{0}),\label{Gamma3_03_rs_expression}\\
\Gamma^{5}_{03}(r_s,y) \approx& -\frac{1}{2}\ln(r_s) + O(r_s^{0}),\label{Gamma5_03_rs_expression}\\
\Gamma^{5}_{04}(r_s,y) \approx& \frac{2\pi}{3\alpha r_s Q(0,y)} + O(r_s^{0}),\label{Gamma5_04_rs_expression}\\
\Gamma^{7}_{04}(r_s,y) \approx& -\frac{1}{2} \ln(r_s) + O(r_s^{0}),\label{Gamma7_04_rs_expression}\\
\Gamma^{3}_{04}(r_s,y) \approx& \frac{4 \pi^2}{3[\alpha r_s Q(0,y)]^2} + O(r_s^{-1}),\label{Gamma3_04_rs_expression}\\
\Gamma^{1}_{04}(r_s,y) \approx& \frac{32 \pi^3}{3 (\alpha r_s Q(0,y))^3} + O(r_s^{-2}).\label{Gamma1_04_rs_expression}
\end{align}

\section{Explicit steps to calculate the integral kind $\mathcal{D}_{n,m}$ and $\beta_n$}
\label{steps_Dnm_beta_n}
In this section, we give the explicit steps to calculate the integrals $\beta_j$ (for $j \geq 1$), $\mathcal{D}_{0,1}$ and $\mathcal{D}_{n,5}$ (for $n=0,2,4,6,8$).
\subsection{Calculating $\beta_n$}
We modify the expression for $\beta_n$ in Eq.~\ref{beta_n} by now defining $\beta_n(a)$ as follows
\begin{equation}
\beta_n(a) = \int^{\infty}_0 dy \frac{1}{a+y^2},
    \label{beta_n(a)}
\end{equation}
where we can calculate $\beta_1(a)$ by doing the change of variable $y=a u$. We obtain
\begin{equation}
\beta_1(a) = \frac{\pi}{2}a^{-\frac{1}{2}}, a > 0.
\label{beta_1_result}    
\end{equation}
By applying the $(n-1)$-th order partial derivative to $\beta_1(a)$ with respect to the variable $a$, we obtain
\begin{align}
\frac{\partial^{n-1}}{\partial a^{n-1}}\beta_1(a) =& (-1)^{n-1} (n-1)!\beta_n(a),
    \label{beta_1_n-1_th_derivative}\\
\frac{\partial^{n-1}}{\partial a^{n-1}}\beta_1(a) =& \frac{\pi}{2}\prod_{i=1}^{n-1}\left(\frac{1}{2}-i\right)a^{-\frac{2n-1}{2}},\label{beta_1_n-1_th_derivative_2}  
\end{align}
where we can solve for $\beta_n(a)$ (for $n>1$) by using these two equations. We obtain
\begin{equation}
\beta_n(a) = \frac{\pi}{2}\frac{1}{(n-1)!}\prod_{i=1}^{n-1}\left(i-\frac{1}{2}\right)a^{-\frac{2n-1}{2}}, \quad a>0. 
    \label{beta_n_a_expression}
\end{equation}
We can make this expression more compact by re-expressing the product of the fractional numbers in terms of factorials. This is performed by completing the factorial by multiplying and dividing by even numbers; we have
\begin{equation}
\prod_{i=1}^{n-1}\left(i-\frac{1}{2}\right) = \frac{(2n-3)!}{2^{2n-3} (n-2)!}, \,\,\mathrm{for}\; n>1.
\label{reorganize_terms3}    
\end{equation}
Using this result, we find $\beta_n(a)$ (for $n\geq 1$) to be
\begin{equation}
\beta_n(a) = \delta_{n,1}\beta_1(a)+\delta_{n>1}\frac{\pi}{2^{2n-2}}\frac{(2n-3)!a^{-\frac{2n-1}{2}}}{(n-1)!(n-2)!},
    \label{beta_n_general}
\end{equation}
where $\delta_{n>1}=1$ for $n>1$ and $\delta_{n>1}=0$ otherwise. The integral kind $\beta_n$ corresponds to the specific case where $a=1$.

\subsection{Calculating some of the integral kind $\mathcal{D}_{0,1}$}
We calculate the integral kind $\mathcal{D}_{0,1}$ starting from its definition
\begin{equation}
\mathcal{D}_{0,1} = \int^{\infty}_0 dy \frac{R(y)}{(1+y^2)},
    \label{D_0,1_expression}
\end{equation}
where we can perform the change of variable $u=1/y$. By doing this, we obtain
\begin{equation}
\mathcal{D}_{0,1} = \int^{\infty}_0 du \frac{u-\tan^{-1}(u)}{u(1+u^2)}.
\label{D_0,1_expression2}    
\end{equation}
Using a partial fraction decomposition, we have
\begin{equation}
\frac{1}{u(1+u^2)} = \frac{1}{u}-\frac{u}{1+u^2},
    \label{partial_fraction_1}
\end{equation}
where the integral of this function is given by
\begin{equation}
\int\frac{du}{u(1+u^2)} = -\frac{1}{2}\ln\left|\frac{1+u^2}{u^2} \right|+C,
\label{integral_partial_fraction_1}    
\end{equation}
where $C$ is a constant of integration. 

Now, integrating by parts in Eq.~\ref{D_0,1_expression2}, we obtain
\begin{align}
\mathcal{D}_{0,1} =& \mathcal{M}_1+\mathcal{M}_2,
\label{D_0,1_expression3}\\
\mathcal{M}_1 =&\frac{1}{2}\int^{\infty}_0 du \ln\left(\frac{1+u^2}{u^2}\right),\label{M_kind_1}\\
\mathcal{M}_2 =& -\frac{1}{2}\int^{\infty}_0 du\ln\left(\frac{1+u^2}{u^2} \right) \frac{1}{1+u^2}.\label{M_kind_2}
\end{align}
We can calculate the integral in Eq.~\ref{M_kind_1} by integration by parts, obtaining
\begin{equation}
\mathcal{M}_1 = \int^{\infty}_0 du \frac{1}{1+u^2} = \frac{\pi}{2}.
\label{M_kind_1_result}    
\end{equation}

Now we proceed to calculate the integral given by Eq.~\ref{M_kind_2} (for $\mathcal{M}_2$). Let's express $\mathcal{M}_2$ as follows
\begin{align}
\mathcal{M}_2 =& \mathcal{M}_{2,1}+\mathcal{M}_{2,2},
\label{M_kind_2_separation}\\
\mathcal{M}_{2,1} =& -\frac{1}{2}\int^{\infty}_0 du \frac{\ln(1+u^2)}{1+u^2},\label{M_2,1_kind}\\
\mathcal{M}_{2,2} =& \int^{\infty}_0 du \frac{\ln(u)}{1+u^2}.\label{M_2,2_kind} 
\end{align}
The integral $\mathcal{M}_{2,1}$ can be evaluated using Feynman's integration trick. Instead of directly calculating $\mathcal{M}_{2,1}$, we evaluate the following parameter-dependent integral
\begin{equation}
\mathcal{M}_{2,1}(a) = -\frac{1}{2}\int^{\infty}_0 du \frac{\ln(1+a u^2)}{1+u^2}.
\label{M_2,1_kind_a}    
\end{equation}
Differentiating with respect to $a$, we obtain
\begin{equation}
\frac{\partial}{\partial a}\mathcal{M}_{2,1}(a) = -\frac{1}{2}\int^{\infty}_0 du \frac{u^2}{(1+u^2)(1+a u^2)}.
\label{M_2,1_kind_a_derivative}    
\end{equation}
Applying a partial fraction decomposition to the integrand, we have
\begin{equation}
\frac{u^2}{(1+u^2)(1+a u^2)} = \frac{1}{a-1}\left[ \frac{1}{1+u^2} -\frac{1}{1+a u^2} \right],
    \label{partial_fraction_2}
\end{equation}
where substituting this expression into Eq.~\ref{M_2,1_kind_a_derivative} yields
\begin{equation}
\frac{\partial}{\partial a} \mathcal{M}_{2,1}(a) = -\frac{\pi}{4\sqrt{a}\left(1+\sqrt{a}\right)}.
\label{M_2,1_kind_a_derivative_result}    
\end{equation}

Now, we proceed to integrate over the $a$ variable in Eq.~\ref{M_2,1_kind_a_derivative_result} from $a=0$ to $a=1$, where $\mathcal{M}_{2,1}(1)=\mathcal{M}_{2,1}$ and $\mathcal{M}_{2,1}(0)=0$. We have
\begin{equation}
\mathcal{M}_{2,1} = -\frac{\pi}{4}\int^{1}_0 da \frac{1}{\sqrt{a}\left(1+\sqrt{a} \right)},
\label{mathcal_M_2,1_kind_integral_a}    
\end{equation}
which can be evaluated by applying the change of variable $b=\sqrt{a}$. We obtain
\begin{equation}
\mathcal{M}_{2,1} = -\frac{1}{2} \int^{1}_0 db \frac{1}{1+b} = -\frac{\ln(2)}{2}.
\label{mathcal_M_2,1_kind_result}    
\end{equation}

\begin{figure*}[htp]
   \begin{center}
   \includegraphics[scale=0.6]{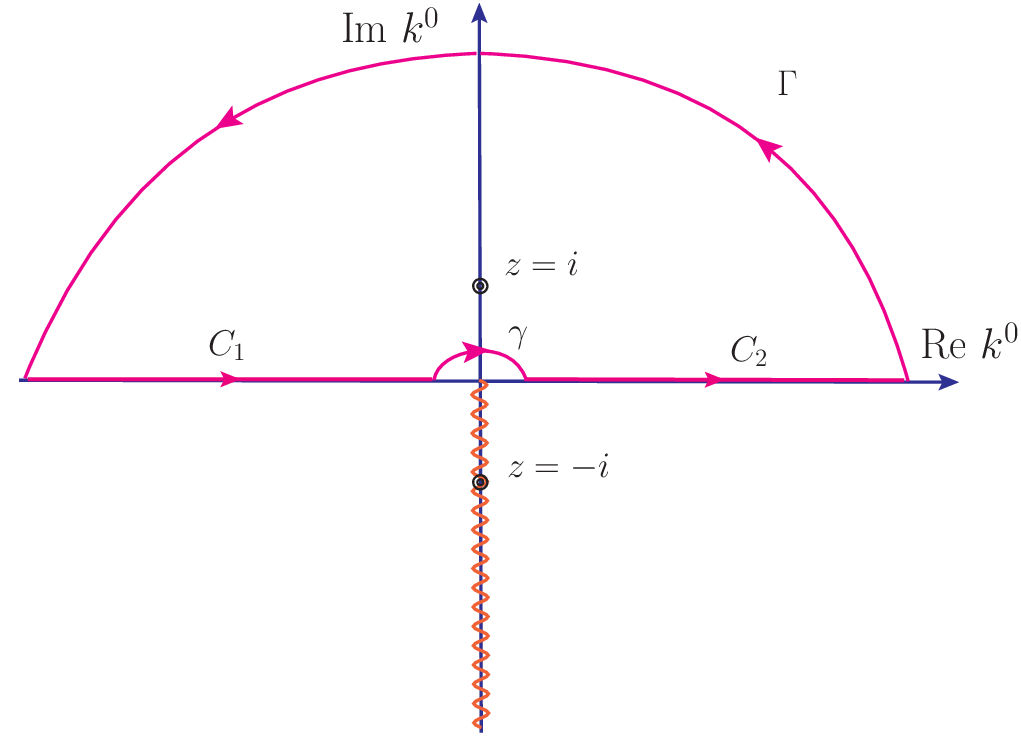} 
   \end{center}
   \caption{contour complex path used for the calculation of the integral $\mathcal{M}_{2,2}$.}
   \label{Cauchy_branch_cut_M22_graph}
\end{figure*}

Now, we proceed to calculate $\mathcal{M}_{2,2}$. This integral can be calculated by using the residue theorem and deforming the complex contour to avoid the branch cut corresponding to the $\ln(u)$ term in the integrand, as shown in Fig.~\ref{Cauchy_branch_cut_M22_graph}. Along path $C$, we have
\begin{equation}
\mathcal{M}_C = \oint_C \frac{\ln(z)}{1+z^2}. 
    \label{integral_complex_C}
\end{equation}

Along the path $\mathcal{C}=C_1+C_2+\gamma+\Gamma$, the integration decomposes as the following sum of integrals
\begin{equation}
\mathcal{M}_{C} = \mathcal{M}_{C_1} + \mathcal{M}_{C_2} + \mathcal{M}_{\gamma} + \mathcal{M}_{\Gamma},
\label{M_C_paths}    
\end{equation}
where the integration along the path $C_2$ yields $\mathcal{M}_{C_2}=\mathcal{M}_{2,2}$. The integrals along the paths $\gamma$ and $\Gamma$ yield zero in the limits $\epsilon \to 0$ and $R \to \infty$, respectively, as proved below. For the path $\gamma$, representing the small keyhole semicircle around the origin, the integral is evaluated by parametrizing $z=\epsilon e^{i \phi}$
\begin{eqnarray}
\mathcal{M}_{\gamma} = -i \epsilon \int^{\pi}_0 d\phi e^{i \phi} \frac{\ln\left(\epsilon e^{i\phi} \right)}{1+\epsilon^2e^{2i\phi}},
\label{integral_gamma_path}
\end{eqnarray}
which yields zero in the limit $\epsilon \to 0$.

The integral along the path $\Gamma$ corresponds to the large semicircle parametrized by $z = R e^{i \phi}$, where the radius is taken to the $R \to \infty$ limit. The integral along this path is given by
\begin{equation}
\mathcal{M}_{\Gamma} = \lim_{R \to \infty} iR\int^{\pi}_0 d\phi e^{i\phi} \frac{\ln(Re^{i\phi})}{1+R^2e^{2i\phi}},
\label{integral_Gamma_path_2}
\end{equation}
which yields zero in the $R \to \infty$ limit.

The integral along the path $C_1$, is given by
\begin{equation}
\mathcal{M}_{C_1} = \lim_{\epsilon \to 0^+}\int^{-\epsilon}_{-\infty} du \frac{\ln(u)}{1+u^2},
\label{integral_C1_path}    
\end{equation}
where performing the change of variable $u=-x$ yields
\begin{equation}
\mathcal{M}_{C_1} = \lim_{\epsilon \to 0^+} \int^{\infty}_0 dx \frac{\ln(-x)}{1+x^2}.
\label{integral_C1_path_change}    
\end{equation}

We note that the logarithm term in Eq.~\ref{integral_C1_path_change} evaluates to
\begin{equation}
\ln(-x) = \ln|x| +i \pi.
\label{logarithm_separation}    
\end{equation}
By substituting Eq.~\ref{logarithm_separation} into Eq.~\ref{integral_C1_path_change}, we obtain
\begin{equation}
\mathcal{M}_{C_1} = \mathcal{M}_{C_2}+i\frac{\pi^2}{2}.
\label{integral_C1_[path_change2]}    
\end{equation}

Applying the residue theorem with the contour avoiding the branch cut in Fig.~\ref{Cauchy_branch_cut_M22_graph}, we find
\begin{equation}
\mathcal{M}_C = i\frac{\pi^2}{2}, 
\label{M_C_residue_theorem}    
\end{equation}
which is obtained by calculating the residue of the integrand associated with $\mathcal{M}_C$ at the simple complex pole $z=i$.

Substituting Eq.~\ref{M_C_residue_theorem} and the evaluated path integrals into Eq.~\ref{M_C_paths}, we obtain
\begin{equation}
\mathcal{M}_{2,2} = 0.
\label{M_2,2_final_expression}    
\end{equation}
Combining the results for $\mathcal{M}_1$, $\mathcal{M}_{2,1}$ and $\mathcal{M}_{2,2}$, we find that $\mathcal{D}_{0,1}$ is given by
\begin{equation}
\mathcal{D}_{0,1} = \frac{\pi}{2}[1-\ln(2)].
 \label{D_0,1_result}   
\end{equation}

\subsection{Calculating some of the integral kind $\mathcal{D}_{n,5}$ (for $n=0,2,4,6,8$)}
Calculating the integral kinds $\mathcal{D}_{n,5}$ (for $n=0,2,4,6,8$), we use the definition given by Eq.~\ref{D_nm_integral}. However, only the terms corresponding to $n=0,2,4,6$ can be handled with the same systematic method used throughout this appendix. For $\mathcal{D}_{8,5}$, we will see that it requires a different treatment.

We begin with the calculation of $\mathcal{D}_{n,5}$ for $n=0,2,4,6,8$. Applying the change of variable $y=1/u$, we obtain
\begin{equation}
\mathcal{D}_{n,5} = \int^{\infty}_0 du \frac{u^{7-n}(u-tan^{-1}(u))}{(1+u^2)^5},
\label{D_n,5,systematic}    
\end{equation}
where $n$ is treated as an even integer ($n=2m$ for $m=0,1,2,3,4$). Consider the integral
\begin{equation}
 \mathcal{A}_m =  \int du \frac{u^{7-2m}}{(1+u^2)^5}, \;\;\mathrm{for}\; m=0,1,2,3,4, 
\label{A_m}    
\end{equation}
which can be solved by doing the change of variable $p=1+u^2$ only for $m=0,1,2,3$. We obtain
\begin{equation}
\mathcal{A}_m = \frac{1}{2}\int dp \frac{(p-1)^{3-m}}{p^5},\;\mathrm{for}\; m=0,1,2,3,
    \label{A_m_2}
\end{equation}
where this integral yields the following result
\begin{equation}
\mathcal{A}_m \!=\! \frac{1}{2}\!\sum^{3-m}_{l=0} \frac{(3-m)!(-1)^{l+m}}{l!(3-m-l)! (4-l)}\frac{1}{(1+u^2)^{4-l}} \!+\!C,
\label{A_m_3}    
\end{equation}
only for $m=0,1,2,3$.

By using Eq.~\ref{A_m_3}, and by using integration by parts in Eq.~\ref{D_n,5,systematic}, we obtain the following integral expression
\begin{align}
\mathcal{D}_{2m,5} =& \frac{1}{2}\sum^{3-m}_{l=0} \frac{(-1)^{m+l+1}}{(4-l)} \nonumber \\
&\times\int^{\infty}_0 \!\!\!du\frac{(3-m)!}{l!(3-m-l)!}\frac{u^2}{(1+u^2)^{5-l}},
\label{D_n,5_systematic_2}    
\end{align}
where this integral expression can be written as a linear combination of the integral kinds $\beta_n$, we have
\begin{align}
\mathcal{D}_{2m,5} =& \frac{1}{2}\sum^{3-m}_{l=0}\frac{(-1)^{m+l+1}}{(4-l)}\nonumber\\
&\times \frac{(3-m)!}{l!(3-m-l)!}\left[\beta_{4-l}-\beta_{5-l} \right],
\label{D_n,5_systematic_3}
\end{align}
where we obtain the values reported in Table~\ref{table_Dnm}, for $\mathcal{D}_{n,m}$ (for $n=0,2,4,6$).

Now we proceed to calculate the integral kind $\mathcal{D}_{8,5}$. By using again the change of variable $y=1/u$, we obtain the following integral
\begin{equation}
\mathcal{D}_{8,5} = \int^{\infty}_0du \frac{u-\tan^{-1}(u)}{u(1+u^2)^5}.
\label{D_8,5_transformed}    
\end{equation}
By performing the separation by partial fractions, we obtain the following
\begin{equation}
\frac{1}{u(1+u^2)^5} = \frac{1}{u} - \sum^{5}_{n=1} \frac{u}{(1+u^2)^n},
\label{partial_fraction_particular}    
\end{equation}
where we can use the expression from Eq.~\ref{partial_fraction_particular} in order to find the following integral result
\begin{equation}
\mathcal{A}_4 = -\frac{1}{2}\ln\left|\frac{1+u^2}{u^2}\right|+ \sum^{4}_{n=1}\frac{1}{2n}\frac{1}{(1+u^2)^n}+C,
\label{integral_partial_fraction_particular}    
\end{equation}
by using Eq.~\ref{integral_partial_fraction_particular}, and integration by parts in Eq.~\ref{D_8,5_transformed}, we obtain the following
\begin{align}
\mathcal{D}_{8,5} =& \mathcal{D}_{0,5}-\sum^{4}_{n=1}\frac{1}{2n}(\beta_{n}-\beta_{n+1}),
\label{D_8,5_transformed_2}\\
\mathcal{D}_{8,5} =& \mathcal{D}_{0,5}-\frac{\beta_1}{2}+\frac{\beta_2}{4}+\frac{\beta_3}{12}+\frac{\beta_4}{24}+\frac{\beta_5}{8},\label{D_8,5_transformed_3}
\end{align}
where by using the tabulated values of $\mathcal{D}_{0,5}$ and the integral kinds $\beta_n$ (for $n=1,2,3,4$), we obtain the value of $\mathcal{D}_{8,5}$ we gave in Table~\ref{table_Dnm}.

\section{Kite-like self-energy series}
\label{Kite_like_self_energy_series}

\begin{figure*}[htp]
   \begin{center}
   \includegraphics[scale=0.37]{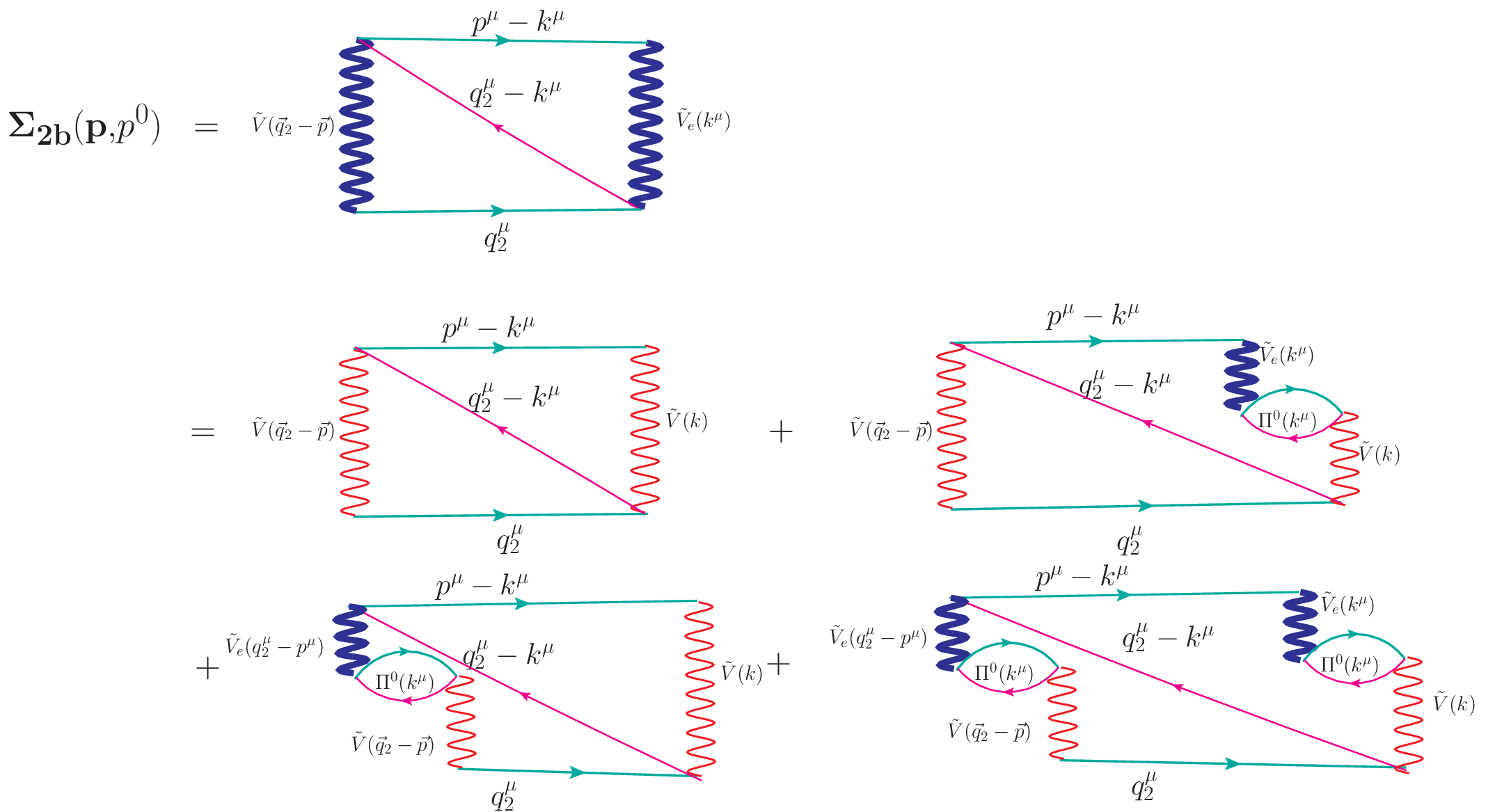} 
   \end{center}
   \caption{Diagrammatic representation of the kite-like self-energy series.}
   \label{self_2b_fig}
\end{figure*}

In this section we prove that the value of the kite-like self-energy series $\Sigma_{2b}(k_F,0)$ evaluated at $p=k_F$ in the static limit yields a constant as a LO contribution in $r_s$ in the high-density limit. This kite-like self-energy series is a second-order term in the RPARPE and can be separated as illustrated in Fig.~\ref{self_2b_fig}, where the first term only depends on the Coulomb interaction $\tilde{V}(q)$ and is the lowest order self-energy diagram from the series, and the second term is what we label as the correction to the lowest order diagram within the renormalized interaction within the RPA. The general expression of $\Sigma_{2b}(p^{\mu})$ is defined as
\begin{align}
\Sigma_{2b}(p^{\mu}) &= \Sigma^0_{2b}(p^{\mu}) + \sum^{3}_{j=1}\Sigma^j_{2b}(p^{\mu}),
\label{kite_diagram_self_energy_definition}\\
\Sigma^0_{2b}(p^{\mu}) &= -\int d[{\bf q_2}]\int d[{\bf k}]A(q_2^{\mu},k^{\mu},p^{\mu})\tilde{V}(\vec{p}-\vec{q}_2)\tilde{V}(k),\label{kite_diagram_self_Onsager}\\
\Sigma^j_{2b}(p^{\mu}) &= -\int d[{\bf q_2}]\int d[{\bf k}]A(q_2^{\mu},k^{\mu},p^{\mu})\tilde{T}_j(p^{\mu},q^{\mu}_2,k^{\mu}),\label{correction_kite_diagram_self}
\end{align}
where $A(q_2^{\mu},k^{\mu},p^{\mu})$ is given
\begin{equation}
A(q_2^{\mu},k^{\mu},p^{\mu}) = G^0(q_2^{\mu})G^0(q_2^{\mu}-k^{\mu})G^0(p^{\mu}-k^{\mu}),
\label{A_function}
\end{equation}
and $\tilde{T}_j(p^{\mu},q^{\mu}_2,k^{\mu})$ are given by the expressions
\begin{align}
\tilde{T}_1(p^{\mu},q^{\mu}_2,k^{\mu}) = -\tilde{V}(\vec{p}-\vec{q}_2)\tilde{V}_i(k^{\mu}),\label{T_1}\\
\tilde{T}_2(p^{\mu},q^{\mu}_2,k^{\mu}) = -\tilde{V}_i(p^{\mu}-q^{\mu}_2)\tilde{V}(k),\label{T_2}\\
\tilde{T}_3(p^{\mu},q^{\mu}_2,k^{\mu}) = \tilde{V}_i(p^{\mu}-q^{\mu}_2)\tilde{V}_i(k^{\mu}),\label{T_3}
\end{align}

The correction to the kite-like self-energy series given by the sum of $\Sigma^j_{2b}(k_F,0)$ for $j=1,2,3$ contributes to higher-order terms in $r_s$ in the high-density limit than $\Sigma^0_{2b}(k_F,0)$, so we only have to focus on this term to extract its LO contribution in $r_s$ to $b_{xc}$. It is straightforward to calculate the integrals over the $q_2^0$ and $k^0$ variables in Eq.~\ref{kite_diagram_self_Onsager}. After performing these integrations and rescaling the momenta into dimensionless variables $\vec{q}_2 \to k_F \vec{q}_2\,'$, $\vec{k} \to k_F \vec{k}\,'$, and $\vec{p} \to k_F \vec{p}\,'$ (with $p\,'^2=1$), we obtain
\begin{align}
\Sigma^0_{2b}(k_F,0) &= \frac{e^4 m}{4 \pi^4}\int d{\bf q_2} \int d{\bf k} \frac{J_2(\vec{q}_2\,',\vec{k}\,',\vec{p}\,')}{D_2(\vec{q}_2\,',\vec{k}\,',\vec{p}\,')},\label{kite_diagram_self_Onsager_2}\\
D_2(\vec{q}_2\,',\vec{k}\,',\vec{p}\,') &= (\vec{q}_2\,'-\vec{p}\,')\cdot \vec{k}\,' |\vec{p}\,'-\vec{q}_2\,'|^2k\,'^2,\label{D2}\\
J_2(\vec{q}_2\,',\vec{k}\,',\vec{p}\,') &= \Theta(1-q_2\,')\Theta(l_1-1)\Theta(l_2-1)\nonumber\\
&+ \Theta(q_2\,'-1)\Theta(1-l_1)\Theta(1-l_2),\label{J2}
\end{align}
where $\vec{l}_1 = \vec{q}_2\,'-\vec{k}\,'$ and $\vec{l}_2 = \vec{p}\,'-\vec{k}\,'$. This proves that $\Sigma_{2b}(k_F,0)$ only contributes as a constant. By substituting the value of this self-energy obtained from Eq.~\ref{kite_diagram_self_Onsager_2} into Eq.~\ref{b_m_final_expression}, we find that the LO contribution in $r_s$ to $b_{xc}$ from $\Sigma_{2b}(k_F,0)$ is of the form $r_s^3$.

\end{document}